\documentclass[final,hidelinks,onefignum,onetabnum]{siamart250211}

\usepackage{lipsum}
\usepackage{amsfonts}
\usepackage{graphicx}
\usepackage{epstopdf}
\usepackage{mathrsfs}
\usepackage{algorithmic}
\usepackage{subcaption}
\ifpdf
  \DeclareGraphicsExtensions{.eps,.pdf,.png,.jpg}
\else
  \DeclareGraphicsExtensions{.eps}
\fi

\newsiamremark{remark}{Remark}
\newsiamremark{hypothesis}{Hypothesis}
\crefname{hypothesis}{Hypothesis}{Hypotheses}
\newsiamthm{claim}{Claim}
\newsiamremark{fact}{Fact}
\crefname{fact}{Fact}{Facts}

\headers{Efficiently Computing the Quasiconcave Envelope}{J. Wu, W. B. Haskell, W. Huang and H. Xu}

\title{Efficiently Computing the Quasiconcave Envelope
with Incomplete Information
}

\author{Jian Wu\thanks{Department of Decision Analytics and Operations, City University of Hong Kong, Hong Kong
  (\email{jwu424@cityu.edu.hk}).}
\and William B. Haskell \thanks{Mitch Daniels School of Business, Purdue University, West Lafayette, Indiana 47907, USA
  (\email{whaskell@purdue.edu}).}
\and Wenjie Huang \thanks{Department of Data and Systems Engineering \& Musketeers Foundation Institute of Data Science, The University of Hong Kong, Pokfulam, Hong Kong
  (\email{huangwj@hku.hk}).}
\and Huifu Xu \thanks{Department of Systems Engineering and Engineering Management, The Chinese University of Hong Kong, Shatin, N.T., Hong Kong
  (\email{hfxu@se.cuhk.edu.hk}).}
  }

\usepackage{amsopn}

\ifpdf
\hypersetup{
  pdftitle={Efficiently Computing the Quasiconcave Envelope 
},
  pdfauthor={J. Wu, W. B. Haskell, W. Huang, and H. Xu}
}
\fi

\begin{document}

\maketitle

\begin{abstract}
In this paper, we study the approximation of an unknown quasiconcave function based on limited partial information.
Available information includes
lower bounds on the values of the target function at a specified set of points, as well as 
some functional properties
including monotonicity, Lipschitz continuity, ranking, and permutation invariance. 
We consider the class of admissible quasiconcave functions that dominate these lower bounds and satisfy 
these 
functional properties.
We then compute the smallest quasiconcave function among the class of admissible quasiconcave functions. 
Specifically, we show how to efficiently compute the quasiconcave envelope (QCoE) of a data sample of points, subject to the additional functional properties. 
The solution procedure takes two steps. 
First, 
a value problem 
is solved
to determine the values of the QCoE on the given data sample. 
Second, an interpolation problem 
is solved to compute the values of the QCoE on other points. Both the value problem and the interpolation problem introduce some theoretical and computational challenges, as they are non-convex and large-scale.
The MILP reformulations of both problems require an exponential number of linear programs (LPs) to be solved in the worst-case.
As our main contribution, we solve the value problem with only a polynomial number of LPs, and
then solve the interpolation problem for any candidate point with only a logarithmic number of LPs. 
Some preliminary numerical 
tests show that the proposed approach 
is efficient and proper.
\end{abstract}



\section{Introduction}

We are motivated by the problem of constructing the quasiconcave envelope (QCoE) of an ambiguous target function based only on limited partial information. We only have some sample data consisting of a finite set of observations of the values of the objective function and some knowledge about its properties.
Based on this information, the QCoE is the minimal quasiconcave function that dominates our data sample and satisfies the desired functional properties.

Let $\Theta = \{ \theta_1, \ldots, \theta_J \} \subset \mathbb{R}^N$ be a finite set of points in the domain of the target function. For each $\theta \in \Theta$, let $\hat{v}(\theta) \in (-\infty, \infty)$ be a lower bound on function values at $\theta$.
We let ${\cal F}$ be the set of all quasiconcave Lipschitz continuous functions $f : \mathbb{R}^N \rightarrow \mathbb{R}$ that dominate this data sample, i.e., they satisfy:
\begin{equation}
\label{eq:majorization}
f(\theta) \geq \hat{v}(\theta),\, \forall \theta \in \Theta.
\end{equation}
The set ${\cal F}$ may also be constrained to satisfy some additional functional properties (like monotonicity and permutation invariance).
We interpret $f \in {\cal F}$ as valuation functions where larger values are preferred.
In addition, we assume ${\cal F}$ contains the true unknown quasiconcave function.

The QCoE of this data sample is denoted by $\psi_{{\cal F}} : \mathbb{R}^N \rightarrow \mathbb{R}$ and is formally defined by:
\begin{equation}
\label{eq:envelope}
\psi_{{\cal F}}(x) = \min_{f \in {\cal F}} f(x),\, \forall x \in \mathbb{R}^N.
\end{equation}
The function $\psi_{{\cal F}}$ returns the point-wise minimum of the admissible functions in ${\cal F}$.
In practical applications, it is known as the worst-case function which gives the lowest possible valuation of a given input.

%
Recall that we stipulate all $f \in {\cal F}$ are Lipschitz continuous, and they satisfy $f(\theta) \geq \hat{v}(\theta) > -\infty$ for at least one $\theta \in \mathbb{R}^N$ by the lower bounds on our sample data. 
Then $\psi_{{\cal F}}(x) > -\infty$ for all $x \in \mathbb{R}^N$ so $\psi_{{\cal F}}$ is well-defined.
The function $\psi_{{\cal F}}$ is itself quasiconcave since quasiconcavity is preserved under minimization.
The minimization problem in Eq.~\eqref{eq:envelope} is both infinite-dimensional and non-convex. Our main objective in this paper is to construct ${\cal F}$ in a principled way and then to efficiently evaluate Eq.~\eqref{eq:envelope} for this ${\cal F}$ for any $x \in \mathbb{R}^N$.


Quasiconcave/quasiconvex functions are ubiquitous in diverse applications.
For instance, the production function in economic models is usually assumed to be quasiconcave \cite{bradley1974fractional,li2019nonparametric,mukherjee2024least} (e.g., the Cobb-Douglas function).
In computer vision problems, the cost function which measures the reprojection errors is often quasiconvex which permits efficient minimization \cite{ke2006uncertainty,ke2007quasiconvex}.
In optimal control (e.g., for wheel-inverted pendulums), the discretized trajectory planning problem minimizes a quasiconvex objective function subject to linear constraints \cite{ning2019time}.

Quasiconcavity/quasiconvexity appear widely in risk management and decision theory.
Quasiconcavity is the most general form of diversification-favoring behavior for functions of random reward.
The certainty equivalent (see, e.g., \cite{Ben-Tal2007}) is a well-known quasiconcave function of random reward.
Cherny and Madan \cite{cherny2009new} and Brown and Sim \cite{brown2009satisficing} develop further quasiconcave functions of random reward.
Quasiconvex risk measures of random costs (which similarly represent diversification-favoring behavior) have also been developed in \cite{mastrogiacomo2015portfolio}.
Brown et al. \cite{brown2012aspirational} present the class of aspirational measures of random reward which are quasiconvex (concentration-seeking) over losses and quasiconcave (diversification-seeking) over gains.

There is extensive research on convex function fitting (see, e.g., \cite{Seijo_2011,lim2012consistency,lim2014convergence}) and constructing convex envelopes (see, e.g., \cite{laraki2008computing,tawarmalani2013explicit}).
The special structure of convex functions means that this class of function fitting problems are often convex optimization problems (see \cite[Section 6.5.5]{Boyd2004}). Armbruster and Delage \cite{armbruster2015decision} develop preference robust optimization (PRO) for random rewards in the expected utility framework. This work essentially constructs the concave envelope of an ambiguous utility function of random rewards based on pairwise comparisons and derivative information.
Delage and Li \cite{delage2017minimizing} carry out a parallel development in PRO for random losses in the convex risk measure framework.
The preceding two works are not strictly function fitting, since they do not take a data sample of function values. Rather, they constrain the values of the ambiguous target function relative to each other (along with a normalization constraint).
Our present paper incorporates both types of information: a data sample of function values and pairwise comparisons between function values.

In contrast to the convex envelope, research on constructing the quasiconcave and/or quasiconvex envelopes is limited.
Chen et al. \cite{chen2021shape} study estimation with shape constraints. They apply functional shape-enforcing operators (for monotonicity, convexity, and quasiconvexity) to obtain functions with the desired properties. They also discuss numerical methods, and enforce the quasiconvexity constraint through bisection search.
Ubhaya \cite{ubhaya1986quasi} studies the problem of approximating a given univariate function with the closest quasiconvex function, based on the supremum norm over an interval.
They propose an algorithm that solves a discrete approximation of this problem on a grid of $n+1$ points with complexity $O(n)$ based on \cite{ubhaya1984n}.
Abbasi and Oberman \cite{abbasi2018computing} compute the quasiconvex envelope of a function, which is equivalent to finding the convex hulls of all level sets. It uses directional line-sweeps and adaptive B-splines to enforce quasiconvexity.
Our work differs from this line of research. We know the ground-truth function is quasiconcave, but we only have limited information about its values on a data sample.

Mukherjee et al.~\cite{mukherjee2024least} study least squares estimation for quasiconvex functions, motivated by the problem in economics of estimating a production function from multiple inputs. They formulate this least squares problem with quasiconvex shape constraints as a mixed integer quadratic program (MIQP). Their formulation depends on interpolation with piecewise constant quasiconvex functions. 
They also develop finite sample bounds and the asymptotic properties of their model, and they identify conditions where the shape-constrained regression is consistent. 

Haskell et al.~\cite{haskell2022preference} develop preference robust optimization (PRO) for quasiconcave choice functions on lotteries (probability distributions).
See, e.g., \cite{frittelli2014risk} which argues why quasiconcavity is a natural property for choice functions on lotteries.
The study in \cite{haskell2022preference} is based on computing the QCoE to approximate a true unknown quasiconcave function defined over the space of lotteries based on partially available information.
They construct the QCoE by solving an MILP, however this MILP quickly becomes intractable for even moderate amounts of preference elicitation data.
In our present paper, we follow this strand of research to consider general QCoEs and quasiconcave interpolation in the space of multivariate functions, which covers a much broader class of problems.
We also include additional partial information about the target function, namely lower bounds on its values on the sample data and the property of permutation invariance.
Permutation invariance in particular makes it much more challenging to construct the QCoE and to do interpolation because it requires exponentially many variables and constraints. Yet, we still obtain efficient algorithms in our present paper.



We now outline the main contributions of
this paper.
We begin by solving a {\it value problem} which determines the values of our quasiconcave function on the given dataset.
The value problem is non-convex (since quasiconcavity is not preserved by convex combination), and its mixed integer linear programming (MILP) reformulation requires an exponential number of variables and constraints.
Yet, we can exploit special structure to solve this problem efficiently with only a polynomial number of LPs of modest size.
Our ``sorting algorithm'' solves the value problem sequentially by computing the value of the QCoE on the sample data one point at a time.
From the solution of the value problem, we can do interpolation to recover the values of the QCoE everywhere else.
The {\it interpolation problem} is similarly non-convex, but we can still solve it based on a logarithmic number of tractable LPs (again, whereas the MILP reformulation requires an exponential number of LPs).
This approach extends the method of constructing the QCoE within the PRO model in \cite{haskell2022preference} to the more general problem of computing the QCoE of a dataset of points. We test our method numerically to estimate an unknown production function. These experiments demonstrate that our sorting algorithm is efficient and scalable, and that interpolation can subsequently be done very rapidly even for large data samples.

The rest of 
the paper is organized as follows.
In Section~\ref{sec:preliminaries}, we review technical preliminaries on quasiconcave functions.
In Section~\ref{sec:envelope}, we formalize the available partial information and the corresponding QCoE.
In Section~\ref{sec:value}, we show how to efficiently solve the value problem with our sorting algorithm.
Then in Section~\ref{sec:interpolation}, we show how to efficiently solve the interpolation problem based on the solution to the value problem.
Section~\ref{sec:extensions} discusses some variations of our main model.
Section~\ref{sec:numerical} reports numerical experiments, and the paper concludes in Section \ref{sec:conclusion}.

\paragraph{Notation}
We let $\mathbb R^N$ denote the set of $N-$dimensional Euclidean vectors (for $N = 1$ we just write $\mathbb R$),
$\langle x, y \rangle$
the scalar 
product 
for $x, y \in \mathbb{R}^N$,
$\|\cdot\|_{1}$ 
the $1-$norm,
and $\|\cdot\|_{\infty}$ 
the supremum norm.
We let $\mathbb R_{\geq 0}^N$ denote the set of vectors in $\mathbb R^N$ with all non-negative components (for $N = 1$ we just write $\mathbb R_{\geq 0}$).
Let $\mathbb N_{\geq 1} := \{1, 2, \ldots \}$ denote the positive integers, and for $I \in \mathbb N_{\geq 1}$ we define the index set $[I] := \{1, 2, \ldots, I\}$.
We let $\textsf{val}(\cdot)$ denote the optimal value of an optimization problem. For $x, y \in \mathbb{R}^N$, we say $x\geq y$ if $x_i \geq y_i$ for all $i \in [N]$

\section{Preliminaries}
\label{sec:preliminaries}

We start by defining some key functional properties.

\begin{definition}
Let $f : \mathbb{R}^N \rightarrow \mathbb{R}$.

(i) [QCo] $f$ is quasiconcave if $f(\lambda x + (1-\lambda)y) \geq \min\{f(x), f(y)\}$ for all $x, y \in \mathbb{R}^N$ and $\lambda \in [0,1]$.


(ii) [Mon] $f$ is monotone if $f(x) \leq f(y)$ for all $x, y \in \mathbb{R}^N$ such that $x \leq y$.

(iii) [Lip] $f$ is $L-$Lipschitz continuous for $L \geq 0$ if $|f(x) - f(y)| \leq L\,\|x-y\|_\infty$ for all $x,\, y \in \mathbb{R}^N$.
\end{definition}
\noindent
We define ${\cal F}_{\text{QCo}}:=\left\{ f : \mathbb{R}^N \rightarrow \mathbb{R} \text{ s.t. [QCo]} \right\}$, ${\cal F}_{\text{Mon}}:=\left\{ f : \mathbb{R}^N \rightarrow \mathbb{R} \text{ s.t. [Mon]} \right\}$, and ${\cal F}_{\text{Lip}}(L):=\left\{ f : \mathbb{R}^N \rightarrow \mathbb{R} \text{ s.t. [Lip]}\right\}$ to be the set of quasiconcave functions, monotone functions, and $L-$Lipschitz continuous functions on $\mathbb{R}^N$, respectively.

For two functions $f, g : \mathbb{R}^N \rightarrow \mathbb{R}$, let $f \wedge g : \mathbb{R}^N \rightarrow \mathbb{R}$ denote the point-wise minimum $(f \wedge g)(x) := \min\{f(x), g(x)\}$ for all $x \in \mathbb{R}^N$. We can extend the definition of $\wedge$ to sets of functions ${\cal F} : \mathbb{R}^N \rightarrow \mathbb{R}$ where $(\wedge {\cal F})(x) := \inf_{f \in {\cal F}} f(x)$ for all $x \in \mathbb{R}^N$.
For any set ${\cal F} \subset {\cal F}_{\text{QCo}}$, $\wedge {\cal F} \in {\cal F}_{\text{QCo}}$ since quasiconcavity is preserved by the infimum.

Next we define an essential class of quasiconcave functions. This class is the key to enforcing the property of quasiconcavity in optimization problems (see \cite{haskell2022preference}).

\begin{definition}
\label{defn:kinked_majorant}
Let $f \in {\cal F}_{\text{QCo}}$, $x \in \mathbb{R}^N$, and $\xi \in \mathbb R^{N}$. Then $h : \mathbb{R}^N \rightarrow \mathbb{R}$ defined by $h(y) := f(x) + \max\{ \langle \xi,\,y-x \rangle,\,0\}$ for all $x \in \mathbb{R}^N$ is a {\em kinked majorant} of $f$ at $x$
if  $h(y) \geq f(y)$ for all $y \in \mathbb{R}^N$.
In that case, $\xi$ is an {\em upper subgradient} of $f$ at $x$. 
The set of all upper subgradients of $f$ at $x$ is called the {\em upper subdifferential} and is denoted by $\partial^+ f(x)$.
\end{definition}
\noindent
Let $h$ be a kinked majorant as in Definition~\ref{defn:kinked_majorant}. Then $h$ has convex upper level sets and is automatically in ${\cal F}_{\text{QCo}}$ by construction. Also, $h$ is $L-$Lipschitz continuous for $L = \|\xi\|_1$. Subsequently, it is well-defined over all of $\mathbb{R}^N$.

The next result characterizes Lipschitz continuous quasiconcave functions with respect to general norms. Let $\|\cdot\|$ be a norm on $\mathbb R^N$, and let $\|\cdot\|_*$ be its dual norm defined by $\|x\|_* := \sup_{\|u\|\leq1} \langle x,u\rangle$ for all $x \in \mathbb R^N$. We specialize this result to the case of $\|\cdot\|_1$ and $\|\cdot\|_{\infty}$ in Theorem~\ref{thm:char-qco-function} below.

\begin{lemma}\label{lem:Lip}
A quasiconcave function $f:\mathbb{R}^N\rightarrow\mathbb{R}$ is $L-$Lipschitz with respect to $\|\cdot\|$ if and only if for every $x\in\mathbb{R}^N$, there exists an upper subgradient $\xi \in \partial^+f(x)$ such that $\|\xi\|_* \leq L$.
\end{lemma}
\begin{proof}
See \cite{plastria1985lower} for the proof of the claim for the Euclidean norm $\|\cdot \|=\|\cdot\|_2$. This proof can be generalized to any norm using the general Cauchy-Schwarz inequality $\langle x,y\rangle\leq \|x\|_{*}\|y\|$ for all $x,\,y \in \mathbb R^N$, as follows. 

Suppose that for every $x\in \mathbb R^{N}$, there exists
an upper subgradient $\xi \in \partial^+f(x)$ 
such that $\|\xi\|_{*}\leq L$. 
Let  $x,y\in \mathbb R^{N}$ and assume WLOG 
that $f(y)\geq f(x)$.
For the case where $f(y)= f(x)$, $f(y)-f(x) \leq L\|y-x\|$ follows trivially.
For the case where $f(y)> f(x)$, 
we have $0<f(y)-f(x) \leq \langle \xi, y-x\rangle\leq \|\xi\|_{*}\|y-x\| \leq L\|y-x\|$. 

For the other direction, suppose that $f$ is $L-$Lipschitz and quasiconcave. For any $x\in \mathbb R^{N}$, let ${\cal S}$ be the upper level set of $f$ at $x$, which is convex by quasiconcavity of $f$. Then, there exists a separating hyperplane for $x$ and ${\cal S}$. Specifically, there exists $\xi_0$ with $\|\xi_0\| = 1$ such that $\langle y-x, \xi_0\rangle\geq 0$ for all $y \in {\cal S}$.

Let $\xi = L\, \xi_0/ \|\xi_0\|_{*}$, we will prove that $\xi$ is an upper-subgradient of $f$. Let $e_0 \in \arg \max_{\|e\|=1}\langle e,\xi\rangle$, then we have $\langle e_0,\xi\rangle= \|\xi\|_{*}$ by definition of the dual norm.
Now fix any $y \in {\cal S}$, and let $y'$ lie on the separating hyperplane (i.e., $\langle \xi_0, y'-x \rangle=0$) such that $y-y' = \|y-y'\|\, e_0$. 
Then, we have 
\begin{align}\label{eq:lipschitz-upper-subgradient-proof-1}
\langle \xi,y-x\rangle = \langle \xi,y-y'\rangle = \langle \xi,e_0\rangle\|y-y'\|= \|\xi\|_{*}\|y-y'\| = L\|y-y'\|.
\end{align}
On the other hand, we have $f(y')\leq f(x)$ since $y'$ lies on the separating hyperplane. By Lipschitz continuity of $f$, it follows that 
\begin{align}\label{eq:lipschitz-upper-subgradient-proof-2}
f(y)-f(x)\leq f(y)-f(y')\leq L\|y-y'\|.
\end{align}
Combining \eqref{eq:lipschitz-upper-subgradient-proof-1} and \eqref{eq:lipschitz-upper-subgradient-proof-2}, we see that $f(y)-f(x)\leq \langle \xi,y-x\rangle$ for all $y\in {\cal S}$. Since $\|\xi\|_{*} = L$ by construction, the desired conclusion follows.
\end{proof}

The next result \cite[Theorem 1]{haskell2022preference} characterizes quasiconcave functions via the class of kinked majorants.

\begin{theorem}\label{thm:char-qco-function}
Let $f:\mathbb{R}^N \rightarrow\mathbb R$. The following assertions hold.

(i) Suppose that $f$ has a kinked majorant at all $x \in \mathbb{R}^N$.
Then, $f$ is quasiconcave, upper semi-continuous, and has a representation
\begin{equation}
f\left(x\right)=\inf_{j\in\mathcal{J}}h_{j}\left(x\right),\,\forall x\in \mathbb{R}^N,\label{eq:Support-quasiconcave}
\end{equation}
where $\mathcal{J}$ is a (possibly infinite) index set and $h_{j}\left(x\right)=\upsilon_j + \max\left\{ \langle \xi_{j},\,x - x_{j}\rangle,\,0\right\}$ for all $j \in \mathcal{J}$ with constants $x_j \in \mathbb{R}^N$, $\upsilon_j \in \mathbb R$, and $\xi_j \in \mathbb R^N$.

(ii) Suppose $f$ is quasiconcave and $L-$Lipschitz continuous with respect to $\|\cdot\|_{\infty}$. Then $f$ has a representation (\ref{eq:Support-quasiconcave}) with $\|\xi_{j}\|_{1}\leq L$ for all $j\in\mathcal{J}$.

(iii) If $f$ has a representation (\ref{eq:Support-quasiconcave}), then
it is quasiconcave.
Moreover, if $\xi_j\geq 0$ for all $j\in\mathcal{J}$, then $f$ is monotone. Conversely,
if $f$ is monotone and quasiconcave, then there exists a set of kinked majorants $\{h_j\}_{j \in {\cal J}}$ with $\xi_j\geq 0$ such that representation (\ref{eq:Support-quasiconcave}) holds.

(iv) For any finite set ${\cal S} \subset \mathbb R^{N}$ and values $\left\{ v(s)\right\}_{s\in{\cal S}}\subset\mathbb R$,
$\hat{f}\text{ : } \mathbb R^{N}\rightarrow\mathbb R$ defined by
\begin{align}
\label{eq:hatfx}
\hat{f}\left(x\right):=\inf_{\upsilon, \xi} \quad & \upsilon
\nonumber
\\
{\rm s.t.} \quad & \upsilon + \max\left\{ \langle \xi,\,s - x\rangle,\,0\right\} \geq v(s),\,\forall s \in {\cal S},\\
& \|\xi\|_{1} \leq L,\nonumber
\end{align}
for all $x \in \mathbb{R}^N$ is quasiconcave. 
Furthermore, the graph of $\hat{f}$ is the (pointwise) minimum of all $L-$Lipschitz quasiconcave majorants
of $\left\{ \left(s,\, v(s)\right)\text{ : }s \in {\cal S} \right\}$.
\end{theorem}

Next we discuss permutation invariant functions. Under permutation invariance, changing the order of some subsets of the components of $x$ does not change the value of $f(x)$. This property is important in applications.
In production functions, changing the order of certain inputs should not change the overall production value. In decision theory, changing the order of potential realizations on different scenarios should not change their evaluation (i.e., law invariance). In social welfare, changing the order of the allocation to certain agents should not change the overall welfare of the group.

We organize the components of $x$ into $M \geq 1$ groups of equal size $K \geq 1$ (so $N = M K$ and $M$ evenly divides $N$).
Then let $\Omega = \{\omega_1, \ldots, \omega_M\}$ index groups, so we can write $x = (x(\omega_1),\ldots,x(\omega_M))$ where $x(\omega_m) := (x_{K (m-1) + 1}, \ldots, x_{K m}) \in \mathbb{R}^K$ for all $m \in [M]$.
For example, suppose we are doing multi-objective stochastic optimization for a vector of random rewards. Then, $K$ would be the number of reward components and $M$ would be the number of uncertain scenarios.

\begin{definition}
(Permutation invariance) (i) Let $\Sigma$ denote the set of all permutations $[M] \rightarrow [M]$, and $\sigma \in \Sigma$ denote a specific permutation. Then $\sigma(x) = (x(\omega_{\sigma(1)}), \ldots, x(\omega_{\sigma(M)}))$ is a permutation of $x$.

(ii) [Prm] $f(x) = f(\sigma(x))$ for all $x \in \mathbb{R}^N$ and $\sigma \in \Sigma$. 
\end{definition}
\noindent
Note that a permutation $\sigma \in \Sigma$ changes the order of the groups of $x$, but not the order of elements within a group. We define ${\cal F}_{\text{Prm}}:=\left\{ f : \mathbb{R}^N \rightarrow \mathbb{R} \text{ s.t. [Prm]} \right\}$ to be the set of all permutation invariant functions.

[Prm] depends on our choice of $M$ and $K$, which gives us flexibility in implementing this property.
In particular, when $K=1$ and $M=N$ then admissible $f$ are permutation invariant over all of their inputs.
When $K=N$ and $M=1$, then permutation invariance is not enforced at all.




\section{Quasiconcave Envelope}
\label{sec:envelope}

In this section, we formalize the problem of constructing the QCoE of a set of points based on available partial information.
We first describe our dataset.
Let $\Theta := \{ \theta_1, \ldots, \theta_J \} \subset \mathbb{R}^N$ be a finite set of $J \geq 1$ inputs.
Each $\theta \in \Theta$ is assigned a lower bound as in Eq.~\eqref{eq:majorization} to form the dataset $\mathsf{D} := \{(\theta, \hat{v}(\theta))\}_{\theta \in \Theta}$. For instance, $\mathsf{D}$ may correspond to the observed points on the graph of a production yield function $f$, and we want to extend $\mathsf{D}$ to obtain a conservative estimate of $f$ on all of $\mathbb{R}^N$. We require admissible $f$ to majorize $\mathsf{D}$ in the following sense.
\begin{itemize}
\item{[Mjr]} (Majorization) $f(\theta) \geq \hat{v}(\theta) > -\infty$ for all $\theta \in \Theta$.
\end{itemize}
\noindent
We define ${\cal F}_{\text{Mjr}}(\mathsf{D}) := \left\{ f : \mathbb{R}^N \rightarrow \mathbb{R} \text{ s.t. [Mjr]} \right\}$.
If $\hat{v}(\theta) = -\infty$, then the constraint $f(\theta) \geq \hat{v}(\theta)$ is vacuous.
The technical condition $\hat{v}(\theta) > -\infty$ for all $\theta \in \Theta$ along with [Lip] ensures that that the QCoE is well-defined. 
Some of the inequalities in [Mjr] may hold with equality for the QCoE. If the data $\mathsf{D}$ come from a quasiconcave function, then they will all hold with equality. However, we have the flexibility to input any set of points and lower bounds (which may not come from a quasiconcave function), and our algorithm will still compute the QCoE without modification.


Let ${\cal R} \subset \Theta \times \Theta$ be a set of pairs of inputs in $\Theta$. The following property constrains the ranking of the values for pairs in ${\cal R}$ (i.e., which one has the larger value). For instance, we may know that one production plan produces at least as much value as another.
\begin{itemize}
\item{[Rnk]} (Ranking) 
$f(\theta) \geq f(\theta')$ for all $(\theta, \theta') \in {\cal R}$.
\end{itemize}
\noindent
We define ${\cal F}_{\text{Rnk}}({\cal R}):=\left\{ f : \mathbb{R}^N \rightarrow \mathbb{R} \text{ s.t. [Rnk]} \right\}$
and the overall set of admissible functions to be:
\begin{equation}
\label{eq:admissible}
{\cal F} = {\cal F}(\mathsf{D}, {\cal R}, L) := {\cal F}_{\text{QCo}} \cap {\cal F}_{\text{Mon}} \cap {\cal F}_{\text{Prm}} \cap {\cal F}_{\text{Mjr}}(\mathsf{D}) \cap {\cal F}_{\text{Rnk}}({\cal R}) \cap {\cal F}_{\text{Lip}}(L),
\end{equation}
where we usually suppress the dependence on $\mathsf{D}$, ${\cal R}$, and $L$.
The set ${\cal F}$ is automatically non-empty, since the constant function $f \equiv \max_{\theta \in \Theta} \hat{v}(\theta)$ satisfies all the conditions of ${\cal F}$.

We seek the QCoE of the data $\mathsf{D}$ and ${\cal R}$ subject to the additional requirements [Mon], [Lip], and [Prm].
The QCoE $\psi_{{\cal F}} : \mathbb{R}^N \rightarrow \mathbb{R}$ is given by the pointwise minimum defined via:
\begin{equation}
\label{eq:psi_F}
     \psi_{{\cal F}}(x) := \inf_{f\in {\cal F}}f(x),\, \forall x \in \mathbb{R}^N.
\end{equation}
Eq.~\eqref{eq:psi_F} can be interpreted as the most conservative estimate of the value of $f(x)$ based on the available partial information.
In the forthcoming discussions, we focus on the QCoE Eq.~\eqref{eq:psi_F} for the specific set of admissible functions in Eq.~\eqref{eq:admissible}.
We have the following basic properties of $\psi_{{\cal F}}$. 

\begin{proposition}
\label{prop:QCoE}
Let $\psi_{{\cal F}}$ be defined as in Eq.~\eqref{eq:psi_F}.

(i) $\psi_{{\cal F}}(x) > -\infty$ for all $x \in \mathbb{R}^N$.

(ii) $\psi_{{\cal F}} \in 
{\cal F}$.
\end{proposition}

\begin{proof}
(i) Since all $f \in {\cal F}$ are $L-$Lipschitz continuous, we have
$$
|\psi_{{\cal F}}(x) - \psi_{{\cal F}}(y)| \leq \sup_{f \in {\cal F}}|f(x) - f(y)| \leq L\,\|x - y\|_{\infty},
$$
using the inequality $|\inf_{f \in {\cal F}} f(x) - \inf_{f \in {\cal F}} f(y)| \leq \sup_{f \in {\cal F}} |f(x) - f(y)|$ for all $x, y \in \mathbb{R}^N$ and [Lip].
For any $\theta \in \Theta$, we have $|\psi_{{\cal F}}(\theta)| < \infty$ by [Mjr]. Then, for any $x \in \mathbb{R}^N$ and some $\theta \in \Theta$, we have $|\psi_{{\cal F}}(x)| \leq |\psi_{{\cal F}}(\theta)| + L\,\|\theta - x\|_{\infty} < \infty$ by [Lip]. It follows that $\psi_{{\cal F}}(x)$ is finite-valued for any $x \in \mathbb{R}^N$.

(ii) Properties [QCo], [Mon], [Lip], [Prm], and [Mjr] are all preserved by $\wedge$.
To verify [Rnk], fix a pair $(\theta, \theta') \in {\cal R}$ and $\epsilon > 0$. Let $f \in {\cal F}$ be such that $\psi_{{\cal F}}(\theta) \geq f(\theta) - \epsilon$. Then we have $ \psi_{{\cal F}}(\theta) \geq f(\theta') - \epsilon \geq \psi_{{\cal F}}(\theta') - \epsilon$ using $f(\theta) \geq f(\theta')$ and $f(\theta') \geq \psi_{{\cal F}}(\theta')$ for all $f \in {\cal F}$. Since $\epsilon$ was arbitrary, the desired result follows.
\end{proof}
\noindent
Without the regularity condition imposed by [Lip], the resulting QCoE $\psi_{{\cal F}}$ will be a discontinuous step function.

Problem~\eqref{eq:psi_F}, which computes $\psi_{{\cal F}}(x)$, is an infinite-dimensional optimization problem.
We need to reformulate Problem~\eqref{eq:psi_F} as a finite-dimensional optimization problem to continue.
Let $\Sigma(\Theta) := \{\sigma(\theta) : \theta \in \Theta, \sigma \in \Sigma\}$ be the set of all permutations of elements of $\Theta$, and let
$$
\widehat{\Theta} := \{ (\theta, \theta') \in \Sigma(\Theta) \times \Sigma(\Theta) : \theta \ne \theta'\},
$$
be the set of all edges in $\Sigma(\Theta)$.
Due to [Prm], we need to assign values to all $\theta \in \Sigma(\Theta)$ (and not just the original $\theta \in \Theta$ in our dataset). Furthermore, to enforce [QCo] under [Prm], we need a majorization constraint for every pair $(\theta, \theta') \in \widehat{\Theta}$.
To enforce [Rnk] under [Prm], we define $\Sigma({\cal R}) := \{(\sigma(\theta), \sigma'(\theta')) : (\theta, \theta') \in {\cal R}, \sigma, \sigma' \in \Sigma\}$ to be the set of all pairs of permutations of $(\theta, \theta') \in {\cal R}$.

For each $\theta \in \Sigma(\Theta)$, let $v(\theta) \in \mathbb{R}$ and $s(\theta) \in \mathbb{R}^N$ be decision variables corresponding to a kinked majorant to $\psi_{{\cal F}}$ at $\theta$. Let $J^{\dagger} := |\Sigma(\Theta)|$ denote the cardinality of $\Sigma(\Theta)$ (which is exponential in $J$). Then, let $v = (v(\theta))_{\theta \in \Sigma(\Theta)} \in \mathbb{R}^{J^{\dagger}}$ and $s = (s(\theta))_{\theta \in \Sigma(\Theta)} \in \mathbb{R}^{J^{\dagger} N}$.
Additionally, let $\upsilon \in \mathbb{R}$ and $\xi \in \mathbb{R}^N$ be decision variables corresponding to a kinked majorant to $\psi_{{\cal F}}$ at $x$.
We then introduce the following disjunctive programming problem to compute $\psi_{{\cal F}}(x)$:
\begin{subequations}\label{prob:psi}
\begin{eqnarray}
(\mathsf{P}(x)) : \min_{v,\,s,\,\upsilon,\,\xi} \, && \upsilon \label{prob:psi-1}\\
{\rm s.t.} \, && v(\theta)+\max \{ \langle s(\theta),\,\theta'-\theta\rangle,\,0 \} \geq v(\theta'), \quad \forall \left(\theta,\,\theta'\right)\in \widehat{\Theta},\label{prob:psi-2}\\
 && v(\theta)\geq v(\theta'), \quad  \forall\left(\theta,\,\theta'\right)\in \Sigma({\cal R}),\label{prob:psi-3}\\
 && v(\sigma(\theta)) \geq \hat{v}(\theta), \quad \forall \theta \in \Theta, \sigma \in \Sigma,\label{prob:psi-4}\\
 && s(\theta)\geq0,\,\|s(\theta)\|_{1}\leq L, \quad\forall \theta\in\Sigma(\Theta),\label{prob:psi-5}\\
 && \upsilon + \max\left\{ \langle \xi,\,\theta - x\rangle,\,0\right\} \geq v(\theta),\,\forall\theta\in\Sigma(\Theta),\label{prob:psi-6}\\
 && \xi \geq 0,\, \|\xi\|_{1} \leq L.\label{prob:psi-7}
\end{eqnarray}
\end{subequations}
The most essential constraint that $f \in {\cal F}_{\text{QCo}}$ is ultimately determined by the values of $f$ on $\Sigma(\Theta)$, reflected in Eq.~\eqref{prob:psi-2}.
The next result establishes correctness of $\mathsf{P}(x)$ for computing $\psi_{{\cal F}}(x)$, it extends \cite[Theorem 2]{haskell2022preference} to our present setting.
\begin{theorem}
\label{thm:psi_reformulation}
For all $x \in \mathbb{R}^N$, we have $\psi_{{\cal F}}(x) = \textsf{val}(\mathsf{P}(x))$ and the optimal value of $\mathsf{P}(x)$ is attained.
\end{theorem}
\begin{proof}
Let $\tilde{\psi}(x; {\cal G}) := \inf_{g \in {\cal G}} g(x)$ be the pointwise infimum for an arbitrary set of functions ${\cal G} : \mathbb{R}^N \rightarrow \mathbb{R}$.
For values $v = (v(\theta))_{\theta \in \Sigma(\Theta)}$, we define
$$
F(v) := \{f : \mathbb{R}^N \rightarrow \mathbb{R} \text{ s.t. } f(\theta) = v(\theta),\, \forall \theta \in \Sigma(\Theta)\}
$$
to be the set of functions that match $v$ on $\Sigma(\Theta)$.
We can first write $\psi_{{\cal F}}(x)$ as an optimization problem over $v$:
\begin{equation*}
     \inf_v \{\tilde{\psi}(x; F(v) \cap {\cal F}) : F(v) \cap {\cal F} \ne \emptyset\},
\end{equation*}
which using the definition of ${\cal F}$ in Eq.~\eqref{eq:admissible} can be expanded to
\begin{eqnarray*}
     \inf_v\, && \tilde{\psi}(x; F(v) \cap {\cal F}_{\text{QCo}} \cap {\cal F}_{\text{Mon}} \cap {\cal F}_{\text{Lip}}(L))\\
     {\rm s.t.}\, && F(v) \cap {\cal F}_{\text{QCo}} \ne \emptyset, F(v) \cap {\cal F}_{\text{Mon}} \ne \emptyset, F(v) \cap {\cal F}_{\text{Lip}}(L) \ne \emptyset, F(v) \cap {\cal F}_{\text{Prm}} \ne \emptyset,\\
     && F(v) \subseteq {\cal F}_{\text{Mjr}}(\mathsf{D}), F(v) \subseteq {\cal F}_{\text{Rnk}}({\cal R}).
\end{eqnarray*}
We enforce the quasiconcavity constraint $F(v) \cap {\cal F}_{\text{QCo}} \ne \emptyset$ by requiring the existence of a kinked majorant at each $\theta \in \Sigma(\Theta)$ in Eq.~\eqref{prob:psi-2} by Theorem~\ref{thm:char-qco-function}(i).
By Theorem~\ref{thm:char-qco-function}(ii), we enforce the Lipschitz continuity constraint $F(v) \cap {\cal F}_{\text{Lip}}(L) \ne \emptyset$ in Eq.~\eqref{prob:psi-5} by requiring $\|s(\theta)\|_1 \leq L$ for all $\theta \in \Sigma(\Theta)$.
By Theorem~\ref{thm:char-qco-function}(iii), we enforce the monotonicity constraint $F(v) \cap {\cal F}_{\text{Mon}} \ne \emptyset$ in Eq.~\eqref{prob:psi-5} by requiring $s(\theta) \geq 0$ for all $\theta \in \Sigma(\Theta)$.
Under [Prm], the constraints $F(v) \subset {\cal F}_{\text{Mjr}}(\mathsf{D})$ are given by Eq.~\eqref{prob:psi-4}.
In addition, the ranking constraints $F(v) \subset {\cal F}_{\text{Rnk}}({\cal R})$ are expressed as $v(\theta) \geq v(\theta')$ for all $(\theta, \theta') \in \Sigma({\cal R})$. This is captured in Eq.~\eqref{prob:psi-3}. Both of these constraints just depend on $v$.

Now we check the constraint $F(v) \cap {\cal F}_{\text{Prm}} \ne \emptyset$ corresponding to [Prm].
Let $(v, s, \upsilon, \xi)$ be a feasible solution of Problem~\eqref{prob:psi}. We will construct a new feasible solution $(\tilde{v}, \tilde{s}, \upsilon, \xi)$ to Problem~\eqref{prob:psi} where $\tilde{v}$ is permutation invariant and $\tilde{v} \leq v$.
For each $\theta \in \Theta$, we implement the following procedure. Let $\bar{\sigma} \in \arg\min_{\sigma \in \Sigma} v(\sigma(\theta))$, then:
\begin{itemize}
    \item Set $\tilde{v}(\sigma(\theta)) = v(\bar{\sigma}(\theta))$ for all $\sigma \in \Sigma$.
    \item Set $\tilde{s}(\sigma(\theta)) = \sigma(\bar{\sigma}^{-1}(s(\bar{\sigma}(\theta))))$ for all $\sigma \in \Sigma$.
\end{itemize}
It follows that $(\tilde{v}, \tilde{s}, \upsilon, \xi)$ is feasible for Problem~\eqref{prob:psi}. To verify this claim for Constraint~\eqref{prob:psi-2}, pick $\theta' \in \Theta$ and $\sigma, \sigma' \in \Sigma$ and compute:
\begin{eqnarray*}
    && \tilde{v}(\sigma(\theta)) + \max\{\langle \tilde{s}(\sigma(\theta)), \sigma'(\theta') - \sigma(\theta) \rangle, 0\}\\
  = && v(\bar{\sigma}(\theta)) + \max\{\langle \sigma(\bar{\sigma}^{-1}(s(\bar{\sigma}(\theta)))), \sigma'(\theta') - \sigma(\theta) \rangle, 0\}\\
  = && v(\bar{\sigma}(\theta)) + \max\{\langle \bar{\sigma}^{-1}(s(\bar{\sigma}(\theta))), \sigma^{-1}(\sigma'(\theta')) - \theta \rangle, 0\}\\
  = && v(\bar{\sigma}(\theta)) + \max\{\langle s(\bar{\sigma}(\theta)), \bar{\sigma}(\sigma^{-1}(\sigma'(\theta'))) - \bar{\sigma}(\theta) \rangle, 0\}\\
  \geq && v(\bar{\sigma}(\sigma^{-1}(\sigma'(\theta'))))\\
  \geq && \tilde{v}(\sigma'(\theta')),
\end{eqnarray*}
where the first inequality follows by feasibility of $(v, s, \upsilon, \xi)$ for Problem~\eqref{prob:psi}, and the second inequality follows by construction of $\tilde{v}$.
Verification for the remaining constraints is immediate.
It follows that for any feasible $(v, s, \upsilon, \xi)$, there exists $(\tilde{v}, \tilde{s}, \upsilon, \xi)$ where $\tilde{v}$ is permutation invariant and $\tilde{v} \leq v$.
Then, without loss of generality we suppose $F(v) \cap {\cal F}_{\text{Prm}} \ne \emptyset$ holds at an optimal solution of $\mathsf{P}(x)$ (since we can always construct such a solution with the above procedure).

Finally, given any feasible $v$, Theorem~\ref{thm:char-qco-function}(iv) characterizes $\tilde{\psi}(x; F(v) \cap {\cal F}_{\text{QCo}} \cap {\cal F}_{\text{Mon}} \cap {\cal F}_{\text{Lip}}(L))$. We conclude that Problem~\eqref{prob:psi} is equivalent to Problem~\eqref{eq:psi_F}.

Now we establish existence of an optimal solution of Problem~\eqref{prob:psi} (and through this equivalence, establish that the optimal value of Problem~\eqref{eq:psi_F} is attained).
We immediately have the lower bounds
$v(\theta) \geq \min_{\theta \in \Theta} \hat{v}(\theta)$ for all $\theta \in \Theta$ by [Mjr], and $\upsilon \geq \min_{\theta \in \Theta} \{ \hat{v}(\theta) - L\|x - \theta\|_1 \}$ by [Mjr] and [Lip].
Next, let $v_{\max} := \max_{\theta \in \Theta} \hat{v}(\theta)$ and let $(v, s, \upsilon, \xi)$ be any feasible solution to Problem~\eqref{prob:psi}. Then let $\tilde v(\theta) := \min\{v(\theta), v_{\max}\}$ and $\tilde s(\theta) := s(\theta)$ for all $\theta \in \Sigma(\Theta)$, $\tilde{\upsilon} := \min\{\upsilon, v_{\max}\}$, and $\tilde{\xi} = \xi$. We see that $(\tilde v, \tilde s, \tilde \upsilon, \tilde \xi)$ is also feasible to Problem~\eqref{prob:psi}, $\tilde \upsilon \leq v_{\max}$, and $\tilde v(\theta) \leq v_{\max}$ for all $\theta \in \Sigma(\Theta)$.
We then obtain the upper bounds $v(\theta) \leq \max_{\theta \in \Theta} \hat{v}(\theta)$ for all $\theta \in \Theta$ and $\upsilon \leq \max_{\theta \in \Theta} \hat{v}(\theta)$.
Combining these lower and upper bounds on $v$ and $\upsilon$, we may add the constraints
\begin{subequations}
\label{eq:v_restricted}
\begin{eqnarray}
v(\theta) \in && \left[ \min_{\theta \in \Theta} \hat{v}(\theta),\, \max_{\theta \in \Theta} \hat{v}(\theta) \right],\, \forall \theta \in \Sigma(\Theta),\\
\upsilon \in && \left[ \min_{\theta \in \Theta} \{\hat{v}(\theta) - L\|x - \theta\|_1\},\, \max_{\theta \in \Theta} \hat{v}(\theta) \right],
\end{eqnarray}
\end{subequations}
to Problem~\eqref{prob:psi} without changing the optimal value or the set of optimal solutions. With the additional Constraint~\eqref{eq:v_restricted}, the feasible region of Problem~\eqref{prob:psi} is compact (since all decision variables are upper and lower bounded, and all of the constraint functions are continuous). The objective of Problem~\eqref{prob:psi} is also continuous, and so this problem has an optimal solution by the Weierstrass Theorem.
\end{proof}

Problem~\eqref{prob:psi} is non-convex due to the constraints Eq.~\eqref{prob:psi-2} and Eq.~\eqref{prob:psi-6} which correspond to [QCo]. Additionally, it has an exponential number of variables and constraints (in $J = |\Theta|$) due to [Prm].
However, we do not solve Problem~\eqref{prob:psi} in its entirety to compute $\psi_{{\cal F}}(x)$ for each new $x \in \mathbb{R}^N$. Instead, we use a two-stage decomposition based on \cite{haskell2022preference}.
Note that if $v$ is fixed in Problem~\eqref{prob:psi}, then we just have to optimize over $(\upsilon, \xi)$ which correspond to a kinked majorant of $\psi_{{\cal F}}$ at $x$.
We first determine the values $\psi_{{\cal F}}$ on $\Sigma(\Theta)$ (this is the value problem described in Section~\ref{sec:value}, and its solution is independent of $x \in \mathbb{R}^N$). Then, we extend to the rest of $\mathbb{R}^N$ (this is the interpolation problem described in Section~\ref{sec:interpolation}, which is different for each new $x \in \mathbb{R}^N$).

\section{Value Problem}
\label{sec:value}

In this section we introduce and solve the value problem which determines the values of $\psi_{{\cal F}}$ on $\Theta$. It can be viewed as a reduction of $\mathsf{P}(x)$ that drops the variables $(\upsilon, \xi)$ corresponding to a kinked majorant of $\psi_{{\cal F}}$ at $x$.
We need a new objective to replace the objective $\upsilon$ of $\mathsf{P}(x)$ which drives all $v$ to be as small as possible. We will show that the sum $\sum_{\theta\in\Sigma(\Theta)} v(\theta)$ is a suitable objective to obtain the value problem:
\begin{subequations}\label{prob:value}
\begin{eqnarray}
(\mathsf{P}) : \min_{v,\,s} \, && \sum_{\theta\in\Sigma(\Theta)}  v(\theta)\label{prob:value-1}\\
{\rm s.t.} \, && v(\theta)+\max \{ \langle s(\theta),\,\theta'-\theta\rangle,\,0 \} \geq v(\theta'), \quad \forall \left(\theta,\,\theta'\right)\in \widehat{\Theta},\label{prob:value-2}\\
 && v(\theta)\geq v(\theta'), \quad  \forall\left(\theta,\,\theta'\right)\in \Sigma({\cal R}),\label{prob:value-3}\\
 && v(\sigma(\theta)) \geq \hat{v}(\theta), \quad \forall \theta \in \Theta, \sigma \in \Sigma,\label{prob:value-4}\\
 && s(\theta)\geq0,\,\|s(\theta)\|_{1}\leq L, \quad\forall \theta\in\Sigma(\Theta).\label{prob:value-5}
\end{eqnarray}
\end{subequations}
%
Problem $\mathsf{P}$ is very similar to $\mathsf{P}(x)$, except the dependence on $x$ is gone and the objective is different.
Problem $\mathsf{P}$ can be thought of as a function fitting problem, where we want the minimal function within a class that dominates our dataset.

The next result establishes that $\mathsf{P}$ correctly determines the values of $\psi_{{\cal F}}$ (defined in Eq.~\eqref{eq:psi_F}) on $\Theta$.
The proof is similar to \cite[Theorem 3]{haskell2022preference}.

\begin{theorem}\label{thm:value}
Let $(v^*, s^*)$ be an optimal solution of Problem~\eqref{prob:value}, then $v^*(\theta) = \psi_{{\cal F}}\left(\theta\right)$ for all $\theta \in \Sigma(\Theta)$.
\end{theorem}
\begin{proof}
For each $\theta \in \Sigma(\Theta)$, by Theorem~\ref{thm:psi_reformulation} we have that $\psi_{{\cal F}}(\theta)$ is given by the optimal value of: 
\begin{subequations}\label{prob:value-theta}
\begin{eqnarray}
(\mathsf{P}(\theta)) : \min_{v,\,s} \, && v(\theta) \label{prob:value-theta-1}\\
{\rm s.t.} \, && v(\theta)+\max \{ \langle s(\theta),\,\theta'-\theta\rangle,\,0 \} \geq v(\theta'), \quad \forall \left(\theta,\,\theta'\right)\in \widehat{\Theta},\label{prob:value-theta-2}\\
 && v(\theta)\geq v(\theta'), \quad  \forall\left(\theta,\,\theta'\right)\in \Sigma({\cal R}),\label{prob:value-theta-3}\\
 && v(\sigma(\theta)) \geq \hat{v}(\theta), \quad \forall \theta \in \Theta, \sigma \in \Sigma,\label{prob:value-theta-4}\\
 && s(\theta)\geq0,\,\|s(\theta)\|_{1}\leq L, \quad\forall \theta\in\Sigma(\Theta).\label{prob:value-theta-5}
\end{eqnarray}
\end{subequations}
An optimal solution $(v^*(\theta'; \theta), s^*(\theta'; \theta))_{\theta' \in \Sigma(\Theta)}$ to Problem~\eqref{prob:value-theta} exists for each $\theta \in \Sigma(\Theta)$ by the same reasoning we used for Problem~\eqref{prob:psi}.

For each $\theta \in \Sigma(\Theta)$, we may define
$$
f_{\theta}^*(x) := \min_{\theta' \in \Sigma(\Theta)}\{v^*(\theta'; \theta) + \max\{\langle s^*(\theta'; \theta), x - \theta' \rangle, 0\}\},\, \forall x \in \mathbb{R}^N,
$$
for which we have $f_{\theta}^* \in {\cal F}$. Next let $g^* := \wedge_{\theta \in \Sigma(\Theta)} f_{\theta}^*$, for which we have $g^* \in {\cal F}$ since ${\cal F}$ is closed under $\wedge$. Finally, let $\tilde{v}(\theta) = g^*(\theta)$ and let $\tilde{s}(\theta) \geq 0$ with $\|\tilde{s}(\theta)\|_1 \leq L$ be an upper subgradient to $g^*$ at $\theta$ (which exists by Theorem~\ref{thm:char-qco-function}(ii)) for all $\theta \in \Sigma(\Theta)$.
It follows that $(\tilde{v}, \tilde{s})$ is feasible for Problem~\eqref{prob:value}.
In addition, $\tilde{v}(\theta) = f_\theta^*(\theta) = v^*(\theta; \theta) = \psi_{{\cal F}}(\theta)$ for all $\theta \in \Sigma(\Theta)$.
Consequently, $(\tilde{v}, \tilde{s})$ is an optimal solution of Problem~\eqref{prob:value}.
\end{proof} 
\noindent
Based on this result, we write $v^*(\theta) = \psi_{{\cal F}}(\theta)$ for all $\theta\in \Sigma(\Theta)$ for the optimal solution of Problem~\eqref{prob:value}.

Problem $\mathsf{P}$ is non-convex due to Constraint~\eqref{prob:value-2}, and it has $O(J^{\dagger})$ decision variables and $O((J^{\dagger})^2)$ constraints (where both are exponential in $J$).
The MILP reformulation of $\mathsf{P}$ has $O((J^{\dagger})^2)$ binary variables and may require $O(2^{(J^{\dagger})^2})$ LPs to be solved in the worst-case.
In its current form, $\mathsf{P}$ is intractable for any meaningfully sized problem.
This motivates us to develop a new sorting algorithm to solve $\mathsf{P}$, which is introduced in the next subsection.
By comparison, our upcoming sorting algorithm has much better theoretical complexity than the MILP reformulation and requires at most $O(J^2)$ LPs to be solved.

\subsection{Sorting Algorithm}

We now describe our sorting algorithm to efficiently solve Problem~\eqref{prob:value} and determine the values of $\psi_{{\cal F}}$ on $\Theta$.
The QCoE $\psi_{{\cal F}}$ naturally induces an ordering over $\Theta$, ranked in descending order according to their $\psi_{{\cal F}}-$values. Our sorting algorithm uses this observation to do the sorting recursively, one input at a time, in descending order of $\psi_{{\cal F}}-$value based on those values that have already been sorted.

Next we define a decomposition of $\Theta$ to be a permutation of $\{(\theta,v^*(\theta))\}_{\theta\in\Theta}$ (the set of pairs of inputs and their $\psi_{{\cal F}}-$values) that is arranged in descending order according to the value of $\psi_{{\cal F}}$. Recall there are $|\Theta| = J$ observations in our data sample.

\begin{definition}
Let each $\theta \in \Theta$ be associated with its $\psi_{{\cal F}}-$value to form the pair $(\theta,v^*(\theta))$.

(i) We call an ordered list $\mathcal{D}:= \left\{(\theta,v^*(\theta))\right\}_{\theta\in\tilde{\Theta}}$ of such pairs a {\em decomposition} of $\Theta$ if: (a) $\tilde{\Theta}$ is a permutation of $\Theta$; and (b) $v^*(\theta) \geq v^*(\theta')$ for all $\theta,\,\theta' \in \tilde{\Theta}$ such that $\theta$ precedes $\theta'$ in $\tilde{\Theta}$.

(ii) For each $t \in [J]$, let $\mathcal{D}_t$ denote the first $t$ elements of $\mathcal{D}$, where $\mathcal{D}_J = \mathcal{D}$.
\end{definition}
\noindent
We treat $\{\mathcal{D}_t\}_{t=1}^{J}$ as ordered lists (rather than sets) because the order of their elements matters. For ordered lists $a$ and $b$, we let $\{a, b\}$ denote the concatenation of $a$ and $b$.
With a slight abuse of notation, \underline{we say that $\theta \in \mathcal{D}_t$ if $(\theta,v^*(\theta)) \in \mathcal{D}_t$}.

To initialize the sorting algorithm, we define
\begin{equation}
\label{eq:initialization}
    \theta_1 \in \arg\max_{\theta \in \Theta} \hat{v}(\theta),\, v^*(\theta_1) = \hat{v}(\theta_1),
\end{equation}
to be the point in our data sample with the largest $\psi_{{\cal F}}-$value (which coincides with the largest lower bound), and then we set $\mathcal{D}_1=\{(\theta_1, v^*(\theta_1))\}$. At the end of each iteration $t \in [J]$, we will have sorted $t$ of the inputs in $\Theta$ to obtain $\mathcal{D}_t$.
Since the decomposition is an ordered list, it induces an ordering on $\Theta$, which coincides with the one induced by $\psi_{{\cal F}}$. Consequently, we recover the indifference curves of $\psi_{{\cal F}}$ where elements with the same $\psi_{{\cal F}}-$value are seated together in $\mathcal{D}$.

\begin{definition}
(i) We let $\underline{v}_t := \min \{v^*(\theta)\mid \theta\in \mathcal{D}_t\}$ denote the smallest $\psi_{{\cal F}}-$value in $\mathcal{D}_t$.

(ii) We say $\theta \in \Theta$ precedes (resp., succeeds) $\theta' \in \Theta$ if $\theta$ precedes $\theta'$ in $\mathcal{D}$ (resp., $\theta$ succeeds $\theta'$ in $\mathcal{D}$). For all $t \in [J-1]$, we say $\theta$ is the successor of $\mathcal{D}_t
\subset {\cal D}$ if $\theta$ is the last element of $\mathcal{D}_{t+1}$, i.e., $\theta\notin \mathcal{D}_t$ but $\theta\in \mathcal{D}_{t+1}$.
\end{definition}

We will construct a decomposition $\mathcal{D}$ according to the following procedure, which relies on an intrinsic feature of kinked majorants. Any kinked majorant of $\psi_{{\cal F}}$ at $\theta$ will always dominate $\theta'$ with smaller $\psi_{{\cal F}}-$values. Thus, we do not need to know the values of such $\theta'$ and can effectively just ignore them when computing the value at $\theta$. This is not possible in the concave case which is based on affine majorants which must dominate the target function everywhere, so we cannot ignore any values.

Problem~\eqref{prob:value} must determine the values on all of $\Sigma(\Theta)$ which has $J^{\dagger}$ elements, so there are exponentially many decision variables. 
In addition, there are exponentially many constraints.
We need to reduce both the number of decision variables and constraints to a polynomial number for practical solution of Problem~\eqref{prob:value} to be possible.
By equivalence with $\psi_{\cal F}$, the optimal solution $v^*$ of Problem~\eqref{prob:value} is permutation invariant, so we just need to determine $J$ values corresponding to $\Theta$ and make sure they are reflected on all of $\Sigma(\Theta)$.
For each $\theta \in \Theta$, the values $v^*(\sigma(\theta))$ must be equal for all $\sigma \in \Sigma$ at the optimal solution of $\mathsf{P}$.
We only need to compute $v^*(\theta)$ for each $\theta\in\Theta$, and then we can assign this same value to all $\Sigma(\theta) := \{\sigma(\theta) : \sigma \in \Sigma\}$.

In each subsequent iteration, given $\mathcal{D}_t$ for $t \in [J-1]$, we want to identify the successor of $\mathcal{D}_t$ to form $\mathcal{D}_{t+1}$.
We only need the information in $\mathcal{D}_t$ to compute  $\psi_{{\cal F}}(\theta)$ for the successor $\theta$ of $\mathcal{D}_t$ (and not the values of any other elements in $\Theta$).
So, given $\mathcal{D}_t$, we want to {\em predict} the value of $\psi_{{\cal F}}(\theta)$ for $\theta \in \Theta \setminus \mathcal{D}_t$ based only on the information in $\mathcal{D}_t$.

For $\mathcal{D}_{t}$ and a candidate $\theta \in \Theta \setminus \mathcal{D}_{t}$, we define the optimization problem:
\begin{subequations}\label{eq:descent-permutationinvariance-multi}
\begin{eqnarray}
(\mathsf{P}_{\textsf{LP}}(\theta;\,\mathcal{D}_{t})) : \min _{\upsilon, \xi} \, && \upsilon \label{eq:descent-permutationinvariance-multi-a}\\
{\rm s.t.} \, && \upsilon+\left\langle \xi, \sigma (\theta')-\theta\right\rangle \geq v^*(\theta'),\,\forall \theta' \in \mathcal{D}_{t},\,\sigma \in \Sigma, \label{eq:descent-permutationinvariance-multi-b}\\
&& \upsilon \geq v^*(\theta'),\, \forall(\theta, \theta') \in {\cal R}, \theta' \in \mathcal{D}_{t}, \label{eq:descent-permutationinvariance-multi-c}\\
&& \upsilon \geq \hat{v}(\theta),\, \xi \geq 0,\, \left\|\xi\right\|_{1} \leq L. \label{eq:descent-permutationinvariance-multi-d}
\end{eqnarray}
\end{subequations}
Note that $\mathsf{P}_{\textsf{LP}}(\theta;\,\mathcal{D}_{t})$ is an LP - it only has affine support function constraints for elements in $\mathcal{D}_t$, and only incorporates information for those elements already sorted in $\mathcal{D}_t$.
Problem $\mathsf{P}_{\textsf{LP}}(\theta;\,\mathcal{D}_{t})$ is part of the value assignment for $\theta$. 
Here $\upsilon$ represents the value of the true unknown quasiconcave function at the point $\theta$, we want it to be as small as possible based on the information in$\mathcal{D}_t$ to determine $\psi_{{\cal F}}(\theta)$. 

Constraint~\eqref{eq:descent-permutationinvariance-multi-b} requires the function $x \rightarrow \upsilon+\left\langle \xi, x -\theta\right\rangle$ to be an affine majorant of the elements of $\mathcal{D}_t$.
In other words, the non-constant part of the kinked majorant of the true unknown quasiconcave function at $\theta$ must majorize its values $v^*(\theta')$ at those points $\theta'\in {\cal D}_t$.
To write this constraint, we also use the fact that $v^*(\sigma(\theta))=v^*(\theta)$ for all $\theta\in \Theta$ and $\sigma\in \Sigma$.

\begin{figure}[ht]
\centering
\begin{subfigure}{0.48\textwidth}
\includegraphics[width=1\linewidth]{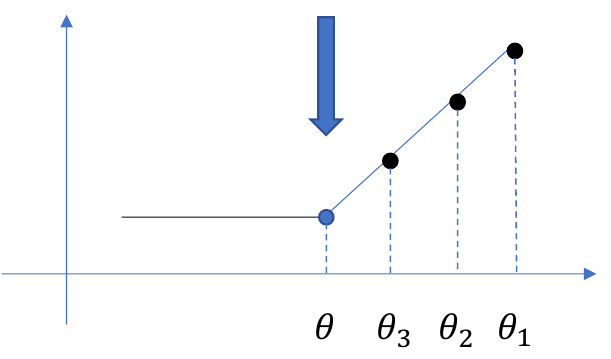} 

\end{subfigure}
\begin{subfigure}{0.48\textwidth}
\includegraphics[width=1\linewidth]{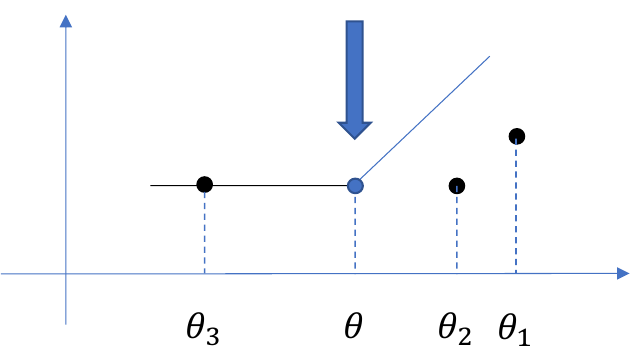}
\end{subfigure}

\caption{Two cases corresponding to the successor $\theta$ of $\mathcal{D}_3$: $\psi_{{\cal F}}(\theta)<\underline{v}_3$ (left) and $\psi_{{\cal F}}(\theta)=\underline{v}_3$ (right)}
\label{fig:single_act_RL}
\end{figure}

Constraint~\eqref{eq:descent-permutationinvariance-multi-c} means that the value assignment at $\theta$ must respect [Rnk] for those inputs that have already been sorted.
Proper reading of Constraint~\eqref{eq:descent-permutationinvariance-multi-b} should be that [Rnk] implies $\upsilon \geq v^*(\theta) \geq v^*(\theta')$ for all $\theta'\in {\cal D}_t$ with $(\theta, \theta') \in {\cal R}$.
Note also that our sorting procedure puts the largest $t$ $\psi_{{\cal F}}-$values of $\{v^*(\theta): \theta\in \Theta\}$ into ${\cal D}_t$ (here we mean 
the function values corresponding to $\theta'\in {\cal D}_t$). However, there could be a case where the $\psi_{{\cal F}}-$value $v^*(\theta)$ at $\theta \in \Theta \setminus \mathcal{D}_t$ is equal to $\underline{v}_t$ (rather than strictly smaller).
In other words, there could be multiple points in $\Theta$ where $\psi_{{\cal F}}$ attains the same value, see Figure \ref{fig:single_act_RL} (right).
Constraint~\eqref{eq:descent-permutationinvariance-multi-c} effectively handles such cases.

Finally, Constraint~\eqref{eq:descent-permutationinvariance-multi-d} means that the value assignment at $\theta$ must respect its own lower bound $\hat{v}(\theta)$ from the dataset $\mathsf{D}$.
In addition, the affine majorant at $\theta$ must respect [Mon] and [Lip].


By definition of a decomposition, we must have $\psi_{{\cal F}}(\theta)\leq \underline{v}_t = \min\{v^*(\theta')\mid \theta' \in \mathcal{D}_t\}$ for all $\theta \in \Theta \setminus \mathcal{D}_t$.
There are thus two cases to consider for the outcome of $\mathsf{P}_{\textsf{LP}}(\theta;\,\mathcal{D}_t)$:
\begin{itemize}
    \item Case one: $\psi_{{\cal F}}(\theta)<\underline{v}_t$ (see Figure \ref{fig:single_act_RL} (left)). In this case, $\mathsf{P}_{\textsf{LP}}(\theta;\,\mathcal{D}_t)$ computes the lowest attainable value of $\psi_{{\cal F}}(\theta)$ given $\mathcal{D}_t$.
    \item Case two: $\psi_{{\cal F}}(\theta)=\underline{v}_t$ (see Figure \ref{fig:single_act_RL} (right)). In this case, the value of $\psi_{{\cal F}}(\theta)$ should just be $\underline{v}_t$. If $\textsf{val}(\mathsf{P}_{\textsf{LP}}(\theta;\,\mathcal{D}_t))\geq \underline{v}_t$ then $\psi_{{\cal F}}(\theta) < \underline{v}_t$ cannot hold and we must have $\psi_{{\cal F}}(\theta) \geq \underline{v}_t$ which gives the equality $\underline{v}_t= \psi_{{\cal F}}(\theta)$ since $\theta \in \Theta \setminus \mathcal{D}_t$.

\end{itemize}

Based on these two cases, we define the function
$$
\pi(\theta;\, \mathcal{D}_t):=\min \left \{\underline{v}_t, \textsf{val}(\mathsf{P}_{\textsf{LP}}(\theta;\,\mathcal{D}_t))\right\},
$$
to be the \textit{predictor} of the value $\psi_{{\cal F}}(\theta)$ for  $\theta \in \Theta \setminus \mathcal{D}_t$.
We will identify the successor of $\mathcal{D}_t$ from those $\theta \in \Theta \setminus {\cal D}_t$ that have the largest predicted value. In particular, the input with the largest predicted value has the largest $\psi_{{\cal F}}-$value among elements not in $\mathcal{D}_t$. Hence this input, along with its predicted value, can be appended to the ordered list $\mathcal{D}_t$ to form $\mathcal{D}_{t+1}$. After that, we find the successor of $\mathcal{D}_{t+1}$ and continue inductively until we compute all of ${\cal D}$. 

Our sorting algorithm for the value problem exactly follows this idea.  In each iteration $t \in [J-1]$, starting with $\mathcal{D}_t$, we compute the predictor $\pi(\theta;\, \mathcal{D}_t)$ for every $\theta \in \Theta \setminus \mathcal{D}_t$ by solving an instance of $\mathsf{P}_{\textsf{LP}}(\theta;\,\mathcal{D}_t)$. Then, some $\theta \in \Theta \setminus \mathcal{D}_t$ that maximizes $\pi(\theta;\, \mathcal{D}_t)$, as well as its predicted value, is appended to $\mathcal{D}_t$. The algorithm terminates with $\mathcal{D} = \mathcal{D}_J$.

We need one more reduction for this sorting algorithm to be computationally practical.
Problem $\mathsf{P}_{\textsf{LP}}(\theta;\,\mathcal{D}_{t})$ has a constraint for every $\theta \in \mathcal{D}_t$ and $\sigma \in \Sigma$, and the cardinality of $\Sigma$ grows exponentially in $J$.
We introduce a reduction of $\mathsf{P}_{\textsf{LP}}(\theta;\,\mathcal{D}_{t})$ as follows to address the exponential number of constraints.
Recall, under [Prm], we can write $x = (x(w_m))^M_{m=1}$ where each $x(w_m) = (x_{K (m-1) + 1}, \ldots, x_{K m}) \in \mathbb{R}^K$.
We define $\vec x_k = (x_k(w_m))^M_{m=1}$ to be the vector consisting of element $k \in [K]$ from each group $x(\omega_m)$ for $m \in [M]$. Then, $\sigma(\vec x_k)=(x_k(\omega_{\sigma(m)}))_{m=1}^M$ is a permutation of $\vec x_k$ which just corresponds to element $k \in [K]$.

We introduce the following additional notation for the tractable reformulation of $\mathsf{P}_{\textsf{LP}}(\theta;\,\mathcal{D}_{t})$.
For $\theta \in \mathbb{R}^N$, we can write $\theta = (\theta(\omega_1),\ldots,\theta(\omega_M))$. Similarly, for $\xi \in \mathbb{R}^N$, we can write $\xi = (\xi(\omega_1),\ldots,\xi(\omega_M))$. Then we define:
\begin{itemize}
    \item $\theta_k = (\theta_k(\omega_1),\ldots,\theta_k(\omega_M)) \in \mathbb{R}^M$ is the vector of values corresponding to element $k \in [K]$ of each of the $M$ groups within $\theta$;
    \item $\xi_k = (\xi_k(\omega_1),\ldots,\xi_k(\omega_M)) \in \mathbb{R}^M$ is the subset of the upper subgradient of $f$ at $\theta$ corresponding to element $k \in [K]$ of each of the $M$ groups of $\xi$;
    \item $y(\theta)\in \mathbb R^M$ and $w(\theta) \in \mathbb R^M$ are auxiliary decision variables for every $\theta\in \mathcal{D}_{t}$. Let $y = (y(\theta'))_{\theta'\in \mathcal{D}_{t}}$ and $w = (w(\theta'))_{\theta'\in \mathcal{D}_{t}}$.
\end{itemize}
We let $\overrightarrow{1} \in \mathbb{R}^M$ denote the vector of all ones, and the transpose operation by $\top$. Our reduced LP is:
\begin{eqnarray*}
(\mathsf{P}_{\textsf{LP}}^{\dagger}(\theta;\,\mathcal{D}_{t})) : \min_{\upsilon, \xi, y, w} \, && \upsilon\\
{\rm s.t.} \, &&  \overrightarrow{1}^{\top} y(\theta')+\overrightarrow{1}^{\top} w(\theta') -\langle \xi, \theta \rangle+ \upsilon-v^*(\theta')\geq 0, \quad \forall \theta'\in \mathcal{D}_{t},\\
&& \sum^K_{k=1}\theta'_{k} \xi_k^{\top}-y(\theta') \overrightarrow{1}^{\top}-\overrightarrow{1} w(\theta')^{\top} \geq 0, \quad \forall \theta'\in \mathcal{D}_{t},\\
&& \upsilon \geq v^*(\theta'), \quad \forall(\theta, \theta') \in {\cal R},\, \theta'\in \mathcal{D}_{t},\\
&& \upsilon \geq \hat{v}(\theta),\, \xi \geq 0,\, \left\|\xi\right\|_{1} \leq L.
\end{eqnarray*}
%
In particular, $\mathsf{P}_{\textsf{LP}}^{\dagger}(\theta;\,\mathcal{D}_{t})$ has a polynomial number of constraints compared to the exponential number of constraints in $\mathsf{P}_{\textsf{LP}}(\theta;\,\mathcal{D}_{t})$.

\begin{proposition}\label{prop:reduction}
For all $t \in [J]$ and $\theta \in \Theta$, $\mathsf{P}_{\textsf{LP}}(\theta;\,\mathcal{D}_{t})$ is equivalent to $\mathsf{P}_{\textsf{LP}}^{\dagger}(\theta;\,\mathcal{D}_{t})$.
\end{proposition}
\begin{proof}
There are exponentially many constraints in \eqref{eq:descent-permutationinvariance-multi-b} due to the index $\sigma \in \Sigma$. Constraint \eqref{eq:descent-permutationinvariance-multi-b} is equivalent to:
\begin{equation}\label{eq:permutation-invariant-exp-constraint}
    \min_{\sigma\in \Sigma} \langle \xi, \sigma (\theta')\rangle -\langle \xi,\theta \rangle+\upsilon-v^*(\theta')\geq 0, \quad \forall \theta'\in \mathcal{D}_{t}. 
\end{equation}
We will reduce the optimization problem $\min_{\sigma\in \Sigma} \langle \xi, \sigma (\theta')\rangle$ in Eq. \eqref{eq:permutation-invariant-exp-constraint}. Recall that $\xi_k$ is the subgradient of $f$ at $\theta$ corresponding to element $k \in [K]$. The optimal value of $\min_{\sigma\in \Sigma} \langle \xi, \sigma (\theta')\rangle$ in Eq. \eqref{eq:permutation-invariant-exp-constraint} is equal to the optimal value of:
\begin{subequations}\label{eq:assignment_multi}
\begin{eqnarray}
\min_{Q \in \mathbb{R}^{M \times M}} \, && \sum^K_{k=1}\xi_k^{\top }Q\, \theta'_{k} \\
{\rm s.t.} \, && Q^\top \overrightarrow{1}=\overrightarrow{1},\\
&& Q \overrightarrow{1}= \overrightarrow{1},\\
&& Q_{m, l} \in\{0,1\}, \quad \forall l,\,m \in [M],
\end{eqnarray}
\end{subequations}
which is a linear assignment problem. Here $Q$ is the permutation matrix corresponding to $\sigma$ so that $Q\, \theta'_k = \sigma(\theta'_k)$ for all $k \in [K]$ (the permutation must be the same for all attributes, hence we only have a single permutation matrix $Q$ in Problem~\eqref{eq:assignment_multi}). Problem~\eqref{eq:assignment_multi} can be solved exactly by relaxing the binary constraints $Q_{m, l} \in\{0,1\}$ to $0 \leq Q_{m, l} \leq 1$ for all $l,\,m \in [M]$. Strong duality holds for the relaxed problem, and the optimal value of the relaxed problem is equal to:
\begin{eqnarray*}
\max_{w,y \in \mathbb R^{M}} \, && \overrightarrow{1}^{\top} w+\overrightarrow{1}^{\top} y \\
{\rm s.t.} \, && \sum^K_{k=1}\theta'_{k} \xi_k^{\top}-w \overrightarrow{1}^{\top}-\overrightarrow{1} y^{\top} \geq 0.
\end{eqnarray*}
It follows that constraint \eqref{eq:permutation-invariant-exp-constraint} is satisfied if and only if there exists $w$ and $y$ such that:
\begin{eqnarray*}
&& \overrightarrow{1}^{\top} w+\overrightarrow{1}^{\top} y -\langle \xi,\theta \rangle+\upsilon-v^*(\theta')\geq 0,\\
&& \sum^K_{k=1}\theta'_{k} \xi_k^{\top}-w \overrightarrow{1}^{\top}-\overrightarrow{1} y^{\top} \geq 0.
\end{eqnarray*}
We substitute the above display into constraint \eqref{eq:descent-permutationinvariance-multi-b} to obtain the desired form.
\end{proof}
\noindent
By Proposition~\ref{prop:reduction}, we have the equivalent form of the predictor given $\mathcal{D}_{t}$:
\begin{equation*}
    \pi(\theta;\, \mathcal{D}_{t}) = \min \{\underline{v}_{t}, \textsf{val}(\mathsf{P}_{\textsf{LP}}^{\dagger}(\theta;\,\mathcal{D}_{t}))\},
\end{equation*}
which is computed for every $\theta \in \Theta \setminus \mathcal{D}_{t}$. Then, some $\theta \in \Theta \setminus \mathcal{D}_{t}$ that maximizes $\pi(\theta;\, \mathcal{D}_{t})$ (as well as its predicted value) is chosen to append to $\mathcal{D}_{t}$, and we proceed to the next iteration.

The details of our sorting algorithm are summarized in Algorithm~\ref{algo:sorting}.
We use the notation $u(\theta) \in \mathbb{R}$ to refer to the value assigned to each $\theta \in \Theta$ in Algorithm~\ref{algo:sorting}.
In the proof of Theorem~\ref{thm:sorting}, we will establish that $u(\theta) = v^*(\theta) = \psi_{{\cal F}}(\theta)$ for all $\theta \in \Theta$, so it assigns the correct values of the QCoE.

\begin{algorithm}
\begin{algorithmic}
\STATE{Initialization: ${\Theta}$, $\mathcal{D}_{1} = \{(\theta_1,v^*(\theta_1))\}$, and $t = 1$;}
\WHILE{$t< J$}
\STATE{Choose $\theta^* \in \arg \max_{\theta \in \Theta \setminus \mathcal{D}_{t}} \pi(\theta;\, \mathcal{D}_{t})$, and set $u(\theta^*): =\pi(\theta^*;\, \mathcal{D}_{t})$;} 

\STATE{Set $\mathcal{D}_{t+1} := \{\mathcal{D}_{t}, (\theta^*, u(\theta^*) )\}$, and set $t := t+1$;}
\ENDWHILE
\RETURN $\mathcal{D}:=\mathcal{D}_{J}$.
 \caption{Sorting algorithm}\label{algo:sorting}
 \end{algorithmic}
\end{algorithm}

The next theorem states that Algorithm~\ref{algo:sorting} returns the values of $\psi_{{\cal F}}$ on $\Theta$ in time polynomial in $J$ (the number of data points in $\mathsf{D}$).

\begin{theorem}\label{thm:sorting}
Algorithm~\ref{algo:sorting} finds a decomposition $\mathcal{D}$ of $\Theta$ and computes $\psi_{{\cal F}}(\theta)$ for all $\theta \in \Sigma(\Theta)$, after solving $O(J^2)$ LPs (where $J = |\Theta|$).
\end{theorem}
\noindent
By Theorem~\ref{thm:sorting}, the computational complexity of Algorithm~\ref{algo:sorting} grows much more slowly in the size of $\mathsf{D}$ (it needs to solve $O(J^2)$ LPs in the worst-case) compared to the MILP reformulation of Problem~\eqref{prob:value} (which needs to solve $O(2^{J^\dagger})$ LPs in the worst-case).

\subsection{Proof of Theorem~\ref{thm:sorting}}

We prove that Algorithm~\ref{algo:sorting} correctly returns a decomposition $\mathcal{D} = \mathcal{D}_{J}$ of $\Theta$ and the value vector $u = (u(\theta))_{\theta\in\Theta}$ corresponds to the values of $\psi_{{\cal F}}$ on $\Theta$. Our proof has three steps:
\begin{enumerate}
    \item First, verify that the candidate solution constructed in Algorithm~\ref{algo:sorting} is well-defined.
    \item Second, verify that this candidate solution gives a lower bound on the optimal solution of $\mathsf{P}$.
    \item Finally, verify that this candidate solution is also feasible for $\mathsf{P}$. Thus, it must be an optimal solution of $\mathsf{P}$.
\end{enumerate}

Our proof is based on a modified version of Algorithm~\ref{algo:sorting} (presented below) that also recursively constructs a candidate solution to $\mathsf{P}$ by selecting one $\theta \in \Theta$ at a time. We will show that this modified algorithm has the properties that we seek, and then we will confirm that it is equivalent to Algorithm~\ref{algo:sorting}. We let $\mathcal{D}_{t}$ (which is composed of tuples $(\theta,u(\theta))$) for $t \in [J-1]$ denote the intermediate sets constructed in the course of this modified algorithm, where we still initialize with $\mathcal{D}_{1} = \{(\theta_1,v^*(\theta_1))\}$.

Given $\mathcal{D}_{t}$, consider the disjunctive program:
\begin{subequations}
\label{eq:20}
\begin{eqnarray}
(\mathsf{P}(\theta;\,\mathcal{D}_{t})) : \min _{\upsilon, \xi} \, && \upsilon\\
{\rm s.t.} \, && \upsilon + \max\{\left\langle \xi, \sigma'(\theta')-\theta\right\rangle,0\} \geq u(\theta'),\,\forall \theta' \in \mathcal{D}_{t},\, \sigma'\in \Sigma,\\
&& \upsilon \geq u(\theta'),\, \forall(\theta, \theta') \in {\cal R}, \theta' \in \mathcal{D}_{t},\\
&& \upsilon \geq \hat{v}(\theta),\, \xi \geq 0,\, \left\|\xi\right\|_{1} \leq L.
\end{eqnarray}
\end{subequations}
Algorithm~\ref{algo:sorting-modified} below is a modified version of Algorithm~\ref{algo:sorting} that solves a sequence of instances of $\mathsf{P}(\theta;\,\mathcal{D}_{t})$.
We do not use Algorithm~\ref{algo:sorting-modified} in actual computation, rather, it is only here to help prove the convergence of Algorithm~\ref{algo:sorting}.
It uses $\textsf{val}(\mathsf{P}(\theta;\,\mathcal{D}_{t}))$ to identify the next $\theta \in \Theta \setminus \mathcal{D}_{t}$ to succeed $\mathcal{D}_t$.
In contrast, Algorithm~\ref{algo:sorting} is based on solving instances of $\mathsf{P}_{\textsf{LP}}(\theta;\,\mathcal{D}_{t})$ (where we actually solve the reduced formulation $\mathsf{P}_{\textsf{LP}}^{\dagger}(\theta;\,\mathcal{D}_{t})$ which is more tractable).
It is more convenient to work with $\mathsf{P}(\theta;\,\mathcal{D}_{t})$ to prove convergence, we later establish equivalence between $\textsf{val}(\mathsf{P}(\theta;\,\mathcal{D}_{t}))$ and $\pi(\theta;\,\mathcal{D}_{t})$.

\begin{algorithm}
\begin{algorithmic}
\STATE{Initialization: ${\Theta}$, $\mathcal{D}_{1} = \{(\theta_1, v^*(\theta_1))\}$, and $t = 1$;}
 \WHILE{$t< J$}
 \STATE{Choose $\theta^* \in \arg \max_{\theta \in \Theta \setminus \mathcal{D}_{t}} \textsf{val}(\mathsf{P}(\theta;\,\mathcal{D}_{t}))$, and set $u(\theta^*):= \textsf{val}(\mathsf{P}(\theta;\,\mathcal{D}_{t}))$;} 
\STATE{Set $\mathcal{D}_{t+1} := \{\mathcal{D}_{t}, (\theta^*, u(\theta^*) )\}$, and set $t := t+1$;}
\ENDWHILE
\RETURN $\mathcal{D}:=\mathcal{D}_{J}$.
 \caption{Modified sorting algorithm} \label{algo:sorting-modified}
 \end{algorithmic}
\end{algorithm}
In iteration $t \in [J-1]$ of Algorithm~\ref{algo:sorting-modified}, we have $\mathcal{D}_{t}$ and select:
$$
\theta \in \arg \max_{\theta' \in \Theta \setminus \mathcal{D}_{t}} \textsf{val}(\mathsf{P}(\theta';\,\mathcal{D}_{t})),
$$
set $u(\theta) := \textsf{val}(\mathsf{P}(\theta;\,\mathcal{D}_{t}))$, and then append $(\theta,u(\theta))$ to $\mathcal{D}_{t}$ to form $\mathcal{D}_{t+1}$.
We verify that this recursive construction of Algorithm~\ref{algo:sorting-modified} is well-defined and monotone.

\begin{lemma}\label{lem:finite-prm}
For all $t \in [J-1]$, $\textsf{val}(\mathsf{P}(\theta;\,\mathcal{D}_{t}))$ is finite.
\end{lemma}
\begin{proof}
For any $x \in \mathbb{R}^N$, $\|\theta - x\|_{\infty} < \infty$ for any $\theta \in \Theta$.
Since $\theta_1$ has finite value $\hat{v}(\theta_1)$, by [Lip] the optimal value $\textsf{val}(\mathsf{P}(x;\,\mathcal{D}_1))$ is finite. By induction, every sorted input in $\mathcal{D}_t$ has finite $u(\theta)-$value and so the optimal value $\textsf{val}(\mathsf{P}(x;\,\mathcal{D}_t))$ is also finite.
Problem~\eqref{eq:20} is a parametric program parameterized by $u(\theta')$ for $\theta'\in {\cal D}_t$.
The conclusion follows directly from the last constraint 
for the lower bound, and the upper bound follows from non-emptiness of the feasible region. 
\end{proof}

\begin{lemma}\label{lem:monotone-prm}
Let $\mathcal{D}$ be the output of Algorithm~\ref{algo:sorting-modified}. Then, $u(\theta)\geq u(\theta')$ if $\theta$ precedes $\theta'$ in $\mathcal{D}$.
\end{lemma}
\begin{proof}
Suppose $\theta_{t+1}$ succeeds $\mathcal{D}_t$ and $\theta_{t+2}$ succeeds $\mathcal{D}_{t+1}$ (which means that $\theta_{t+2}$ must succeed $\theta_{t+1}$ in $\mathcal{D}$). Let $(\upsilon, \xi)$ be an optimal solution of $\mathsf{P}(\theta_{t+2};\,\mathcal{D}_{t})$. By definition of Algorithm~\ref{algo:sorting-modified} we have:
$$
\theta_{t+1} \in \arg \max_{ \theta \in \Theta \setminus \mathcal{D}_t} \textsf{val}(\mathsf{P}( \theta;\,\mathcal{D}_t)),
$$
and so it follows that $u(\theta_{t+1}) = \textsf{val}(\mathsf{P}(\theta_{t+1};\,\mathcal{D}_t))\geq \textsf{val}(\mathsf{P}(\theta_{t+2};\,\mathcal{D}_t)) = \upsilon$.
Next, notice that the only additional constraints in $\mathsf{P}(\theta_{t+2};\,\mathcal{D}_{t+1})$ over $\mathsf{P}(\theta_{t+2};\,\mathcal{D}_{t})$ are:
\begin{subequations}
\label{eq:modified_monotone}
\begin{eqnarray}
&& \upsilon + \max\{\left\langle \xi, \sigma'(\theta_{t+1})-\theta_{t+2}\right\rangle,0\} \geq u(\theta_{t+1}),\,\forall \sigma'\in \Sigma,\\
&& \upsilon \geq u(\theta_{t+1}),\, \forall (\theta, \theta_{t+1}) \in {\cal R}.
\end{eqnarray}
\end{subequations}
Since $(\upsilon,\xi)$ is feasible for $\mathsf{P}(\theta_{t+2};\,\mathcal{D}_{t})$, $u(\theta_{t+1}) \geq \upsilon$, and Eq.~\eqref{eq:modified_monotone} is satisfied by $\upsilon = u(\theta_{t+1})$ for any $\xi$, we see that $(u(\theta_{t+1}),\xi)$ is feasible for $\mathsf{P}(\theta_{t+2};\,\mathcal{D}_{t+1})$.
Thus, we have $u(\theta_{t+2}) = \textsf{val}(\mathsf{P}(\theta_{t+2};\,\mathcal{D}_{t+1})) \leq u(\theta_{t+1})$, and the conclusion follows by induction.
\end{proof}

Next we will verify that Algorithm~\ref{algo:sorting-modified} produces a lower bound on the optimal value of $\mathsf{P}$. For any $\theta \in \Theta$, we recall $\mathsf{P}(\theta)$ just minimizes the value at $\theta \in \Theta$. Problem $\mathsf{P}(\theta)$ has the finite reformulation given in Eqs.~\eqref{prob:value-theta-1} - \eqref{prob:value-theta-5}.

\begin{lemma}\label{lem:lower_bound-prm}
For all $t \in [J-1]$, $\textsf{val}(\mathsf{P}(\theta;\,\mathcal{D}_{t})) \leq \psi_{{\cal F}}(\theta)$ for all $\theta \notin \mathcal{D}_{t}$.
\end{lemma}
\begin{proof}
We prove this statement by induction starting with $\mathcal{D}_{1} = \{(\theta_1, v^*(\theta_1))\}$. We see $\textsf{val}(\mathsf{P}(\theta;\,\mathcal{D}_{1}))$ is a lower bound for $\psi_{{\cal F}}(\theta)$ for all $\theta \notin \mathcal{D}_{1}$ (this is automatic because $\mathsf{P}(\theta;\,\mathcal{D}_{1}))$ has fewer constraints than Problem \eqref{prob:value-theta} which computes $\psi_{{\cal F}}(\theta)$ exactly). Since $\psi_{{\cal F}}(\theta) = \psi_{{\cal F}}(\sigma(\theta))$ for all $\sigma\in\Sigma$, $\textsf{val}(\mathsf{P}(\theta;\,\mathcal{D}_{1}))$ is also a lower bound for $\psi_{{\cal F}}(\sigma(\theta))$. 
Proceeding inductively, if every estimate in $\mathcal{D}_{t}$ is a lower bound on the corresponding value $\psi_{{\cal F}}(\theta')$ for all $\theta' \in \mathcal{D}_{t}$, then the optimal value of $\mathsf{P}(\theta;\,\mathcal{D}_{t})$ is a lower bound for all $\theta \notin \mathcal{D}_{t}$ (this follows because $\mathsf{P}(\theta;\,\mathcal{D}_{t})$ has fewer constraints than Problem \eqref{prob:value-theta}, and by induction all of the values $v(\theta')$ for $\theta' \in \mathcal{D}_{t}$ are themselves lower bounds for $\psi_{{\cal F}}(\theta')$).
 \end{proof}

Next, we will verify that the construction of Algorithm~\ref{algo:sorting-modified} is also feasible for $\mathsf{P}$.
\begin{lemma}\label{lem:feasible-prm}
Let $v=(v(\sigma(\theta)))_{\theta \in \Theta,\,\sigma \in \Sigma}$ where $v(\sigma(\theta))=u(\theta)$ for all $\sigma \in \Sigma$ and $\theta\in \Theta$. Then there exists $(v, s)$ that is feasible for $\mathsf{P}$.
\end{lemma}
\begin{proof}
Suppose $\theta \in \Theta$ succeeds $\mathcal{D}_{t}$ for some $t$. By Lemma \ref{lem:monotone-prm}, we have $u(\theta) \geq u(\theta')$ for all $\theta'\in \Theta \setminus \mathcal{D}_{t}$. As a result, Eq. \eqref{prob:value-2} is satisfied for all pairs $(\theta,\theta')$ with $\theta' \in \Theta \setminus \mathcal{D}_{t}$ and $\sigma,\sigma'\in \Sigma$, and hence for all $(\theta,\theta')$ with $\theta'\in \Theta$ and $\sigma,\sigma'\in \Sigma$ (because this constraint is already satisfied for all $(\theta,\theta')$ with $\theta' \in\mathcal{D}_{t}$ and $\sigma,\sigma'\in \Sigma$ in $\mathsf{P}(\theta;\mathcal{D}_{t})$). Constraint \eqref{prob:value-3} is also satisfied for all $(\theta,\theta')$ with $\theta' \in \Theta \setminus \mathcal{D}_{t}$ and $\sigma,\sigma'\in \Sigma$, and hence for all $(\theta,\theta')$ with $\theta' \in\Theta$ and $\sigma,\sigma'\in \Sigma$ (again, because this constraint is already satisfied for all $(\theta,\theta')$ with $\theta' \in\mathcal{D}_{t}$ and $\sigma,\sigma'\in \Sigma$ in $\mathsf{P}(\theta;\mathcal{D}_{t})$). This reasoning applies to every $\theta\in\Theta$, and so the constructed candidate solution is feasible. 
\end{proof}

Now to complete our proof of the correctness of the modified sorting algorithm, we will establish $\textsf{val}(\mathsf{P}(\theta;\,\mathcal{D}_{t})) = \min\{ \textsf{val}(\mathsf{P}_{\textsf{LP}}(\theta;\,\mathcal{D}_{t})),\, \underline{v}_{t}\} = \pi(\theta;\mathcal{D}_{t})$.

\begin{lemma}\label{lem:lp-prm}
For all $t \in [J-1]$, $\textsf{val}(\mathsf{P}(\theta;\,\mathcal{D}_{t})) = \pi(\theta;\mathcal{D}_{t})$ for all $\theta \notin \mathcal{D}_{t}$.
\end{lemma}
\begin{proof}
We have shown $u(\sigma(\theta)) = u(\sigma'(\theta))$ for all $\theta$ and $\sigma,\sigma'\in \Sigma$.
Next, we note that every feasible solution of $\mathsf{P}_{\textsf{LP}}(\theta;\, \mathcal{D}_t)$ is also feasible for $\mathsf{P}(\theta;\, \mathcal{D}_t)$, so we must have $\textsf{val}(\mathsf{P}(\theta;\, \mathcal{D}_t)) \leq \textsf{val}(\mathsf{P}_{\textsf{LP}}(\theta;\, \mathcal{D}_t))$. Now let $\upsilon = \textsf{val}(\mathsf{P}(\theta;\, \mathcal{D}_t))$ and consider the two cases: (i) $\upsilon<\underline{v}_t$ and (ii) $\upsilon= \underline{v}_t$ (due to monotonicity, the case $\upsilon>\underline{v}_t$ will not occur).

In the first case, we have $\upsilon< u(\theta')$ for every $\theta'\in \mathcal{D}_t$. Thus, the constraint
$$
\upsilon+\max\{\left\langle \xi, \theta'-\theta\right\rangle,0\} \geq u(\theta'),
$$
in $\mathsf{P}(\theta;\,\mathcal{D}_t)$ is equivalent to $\upsilon+\left\langle \xi, \theta'-\theta\right\rangle \geq u(\theta')$ (or otherwise it would be infeasible). Hence, $\mathsf{P}(\theta;\,\mathcal{D}_t)$ is equivalent to $\mathsf{P}_{\textsf{LP}}(\theta;\,\mathcal{D}_t)$ and we have $\upsilon = \textsf{val}(\mathsf{P}(\theta;\,\mathcal{D}_t))$.

In the second case, we have $\underline{v}_t= \upsilon = \textsf{val}(\mathsf{P}(\theta;\,\mathcal{D}_t)) \leq \textsf{val}(\mathsf{P}_{\textsf{LP}}(\theta;\,\mathcal{D}_t))$, and it follows that $\upsilon = \underline{v}_t = \pi(\theta;\,\mathcal{D}_t)$. To conclude, in both cases we have $\upsilon=\pi(\theta;\,\mathcal{D}_t)$.
\end{proof}

By Lemma \ref{lem:lp-prm}, Algorithm~\ref{algo:sorting} and Algorithm~\ref{algo:sorting-modified} are equivalent. Since the output of Algorithm~\ref{algo:sorting} is both a lower bound and a feasible solution, it follows that Algorithm~\ref{algo:sorting} returns an optimal solution of $\mathsf{P}$, completing the proof of Theorem~\ref{thm:sorting}.

\section{Interpolation}
\label{sec:interpolation}

An optimal solution to $\mathsf{P}$ yields the values of $\psi_{{\cal F}}$ on $\Theta$ by Theorem~\ref{thm:value}.
Now we explain how to compute $\psi_{{\cal F}}(x)$ for general $x \in \mathbb{R}^N$ interpolating the values $v^*$ of the QCoE on $\Theta$.
We will show that $\psi_{{\cal F}}(x)$ is given by the optimal value of:
\begin{subequations}\label{prob:interpolation}
\begin{eqnarray}
(\mathsf{P}(x;\,v^*)) : \min_{\upsilon,\, \xi} \, && \upsilon \\
{\rm s.t.} \, && \upsilon + \max\{ \langle \xi,\,\sigma(\theta)-x \rangle,\,0\} \geq v^*(\theta), \quad \forall \theta\in\Theta,\,\sigma \in \Sigma,\label{prob:interpolation-1}\\
 && \xi \geq 0,\,\|\xi\|_{1}\leq L.\label{prob:interpolation-2}
\end{eqnarray}
\end{subequations}
To do the interpolation, we want to find the smallest possible value for $f(x)$ among functions $f \in {\cal F}_{\text{QCo}} \cap {\cal F}_{\text{Mon}} \cap {\cal F}_{\text{Lip}}(L)$ that dominate the points $\{(\theta, v^*(\theta))\}_{\theta \in \Sigma(\Theta)}$.
By Theorem~\ref{thm:char-qco-function}, we may restrict $f$ to the class of kinked majorants in $\mathsf{P}(x;\,v^*)$.
Eq.~\eqref{prob:interpolation-1} ensures that the kinked majorant $y \rightarrow \max \{ \langle \xi,\, y - x \rangle,\, 0 \}$ majorizes $\{(\theta, v^*(\theta))\}_{\theta \in \Sigma(\Theta)}$, while Eq.~\eqref{prob:interpolation-2} enforces [Mon] and [Lip]. 
Properties [Mjr] and [Rnk] are already accounted for by $v^*$.
The next theorem 
states that
$\psi_{{\cal F}}(x)$ 
can be obtained
by solving the interpolation problem $\mathsf{P}(x; v^*)$.

\begin{theorem}\label{thm:interpolation}
We have $\psi_{{\cal F}}(x) = \textsf{val}(\mathsf{P}(x;\,v^*))$ for all $x \in \mathbb{R}^N$.
\end{theorem}
\begin{proof}
First we define the function $\tilde \psi : \mathbb{R}^N \rightarrow \mathbb{R}$ via:
\begin{eqnarray*}
\tilde{\psi}(x) := \inf_f\, && f(x)\\
{\rm s.t.}\, && f(\theta) \geq \psi_{{\cal F}}(\theta),\,\forall \theta \in \Sigma(\Theta),\\
&& f \in {\cal F}_{\text{QCo}} \cap {\cal F}_{\text{Mon}} \cap {\cal F}_{\text{Lip}}(L),
\end{eqnarray*}
for all $x \in \mathbb{R}^N$.
Since $\psi_{{\cal F}} \in {\cal F}_{\text{QCo}} \cap {\cal F}_{\text{Mon}} \cap {\cal F}_{\text{Lip}}(L)$, we have $\tilde{\psi}(\theta) = \psi_{{\cal F}}(\theta)$ for all $\theta \in \Sigma(\Theta)$.
It is then immediate that $\tilde{\psi} \leq \psi_{{\cal F}}$, since $\tilde{\psi}$ is only required to majorize $\psi_{{\cal F}}$ on $\Sigma(\Theta)$ (and not on all of $\mathbb{R}^N$).

For the other direction, first note $\tilde{\psi} \in {\cal F}_{\text{QCo}} \cap {\cal F}_{\text{Mon}} \cap {\cal F}_{\text{Lip}}(L)$ since these properties are all preserved by the infimum.
Since $\tilde{\psi}(\theta) = \psi_{{\cal F}}(\theta)$ for all $\theta \in \Sigma(\Theta)$, we have $\tilde{\psi} \in {\cal F}_{\text{Mjr}}(\mathsf{D}) \cap {\cal F}_{\text{Rnk}}({\cal R})$ because these properties only depend on the values on $\Sigma(\Theta)$.
Now we will show that $\tilde{\psi}$ is permutation invariant, i.e., $\tilde{\psi}(\sigma(x)) = \tilde{\psi}(\sigma'(x))$ for all $\sigma, \sigma' \in \Sigma$.
By Theorem~\ref{thm:char-qco-function}(iv), $\tilde{\psi}(x)$ has the equivalent form $\tilde{\psi}(x) = \textsf{val}(\mathsf{P}(x;\,v^*))$.
Let $(\upsilon, \xi)$ be a feasible solution to $(\mathsf{P}(\sigma(x);\,v^*))$ for some $\sigma \in \Sigma$, and then let $\upsilon' = \upsilon$ and $\xi' = \sigma'(\sigma^{-1}(\xi))$ for some $\sigma' \in \Sigma$. To see that $(\upsilon', \xi')$ is feasible for $(\mathsf{P}(\sigma'(x);\,v^*))$, we have:
\begin{eqnarray*}
    && \upsilon' + \max\{\langle \xi', \sigma''(\theta) - \sigma'(x)\}\\
    = && \upsilon + \max\{\langle \sigma'(\sigma^{-1}(\xi)), \sigma''(\theta) - \sigma'(x)\}\\
    = && \upsilon + \max\{\langle \xi, \sigma((\sigma')^{-1}(\sigma''(\theta))) - \sigma(x)\} \geq \psi_{{\cal F}}(\theta),
\end{eqnarray*}
for all $\theta \in \Theta$ and $\sigma'' \in \Sigma$.
We conclude that the value of $\tilde{\psi}(\sigma(x))$ is constant for all $\sigma \in \Sigma$, so $\tilde{\psi}$ satisfies [Prm] and $\tilde{\psi} \in {\cal F}$. It follows that $\psi_{{\cal F}} \leq \tilde{\psi}$ must also hold, showing that $\psi_{{\cal F}} = \tilde{\psi}$.
\end{proof}

Problem~\eqref{prob:interpolation} is non-convex and has an exponential number of constraints due to Eq.~\eqref{prob:interpolation-1}. 
Instead of trying to solve it directly, we can apply Algorithm~\ref{algo:sorting} to the augmented set of inputs $\Theta \cup \{x\}$ to determine $\psi_{{\cal F}}(x)$. This is exactly what we would do if $x$ were an input in our original dataset $\mathsf{D}$. However, we are only trying to determine the value of $\psi_{{\cal F}}(x)$ so we can stop as soon as $x$ has been sorted.

Recall $J$ is the number of data points in $\mathsf{D}$. We can search over $[J]$ to see where $x$ should be inserted into the existing decomposition $\mathcal{D}$.
Specifically, we can evaluate $\pi(x;\, \mathcal{D}_t)$ (based on solving $\mathsf{P}^{\dagger}_{\textsf{LP}}(x;\,\mathcal{D}_t)$) for different $t \in [J]$, and then compare these values to $v^*$ to determine where $x$ belongs in the augmented decomposition of $\Theta \cup \{x\}$.
In fact, we can do binary search over $[J]$ to compute $\psi_{{\cal F}}(x)$ by solving at most $O(\log J)$ LPs, see Algorithm~\ref{algo:interpolation-binary}.
We define $\underline{v}_{J+1} := -\infty$ to allow Algorithm~\ref{algo:interpolation-binary} to check for values lower than $\underline{v}_{J}$.

\begin{algorithm} 
\begin{algorithmic}
\STATE{Initialization: $v^*$, $x$, $t_{\text{upper}}=1$, and $t_{\text{lower}}=J$;}
 \WHILE{$t_{\text{lower}} \neq t_{\text{upper}}$}
 \STATE{Set $t:=\lfloor (t_{\text{upper}} + t_{\text{lower}})/2 \rfloor$;}
\IF{$\pi(x;\, \mathcal{D}_t) < \underline{v}_{t+1}$,} 
\STATE{set $t_{\text{upper}}:=t+1$;} 
\ELSE
\STATE{set $t_{\text{lower}}:=t$;}
\ENDIF
\ENDWHILE
\STATE{Set $t:=\lfloor (t_{\text{upper}} + t_{\text{lower}})/2 \rfloor$;}
\RETURN $\psi_{{\cal F}}(x) = \pi(x;\, \mathcal{D}_t)$.
 \caption{Interpolation algorithm}\label{algo:interpolation-binary}
 \end{algorithmic}
\end{algorithm}

Recall that the elements of ${\cal D}_J$ are listed in decreasing order (so lower indices correspond to larger values of $\psi_{{\cal F}}$).
In Algorithm~\ref{algo:interpolation-binary}, $t_{\text{upper}}$ represents the upper end-point and $t_{\text{lower}}$ represents the lower end-point of the interval we are currently searching. Recall $\lfloor \cdot \rfloor$ is the floor function where $\lfloor t' \rfloor$ returns the largest integer $t$ such that $t \leq t'$.
Then, $\lfloor (t_{\text{upper}} + t_{\text{lower}})/2 \rfloor$ serves as the midpoint of the interval we are currently searching.

\begin{theorem}
\label{thm:interpolation-binary}
Fix $x \in \mathbb{R}^N$, Algorithm~\ref{algo:interpolation-binary} returns $\psi_{{\cal F}}(x)$ after solving at most $O(\log J)$ LPs (where $J = |\Theta|$).
\end{theorem}
\begin{proof}
Imagine applying Algorithm~\ref{algo:sorting} to the augmented set $\Theta \cup \{x\}$, where we set $\hat{v}(x) = -\infty$ so there is no a priori lower bound on function values at $x$.
If $\underline{v}_{t+1} < \pi(x;\, \mathcal{D}_t) \leq \underline{v}_t$, then by Lemma~\ref{lem:lp-prm} we have $\psi_{{\cal F}}(x) = \pi(\theta;\mathcal{D}_{t})$.
There is a unique $t$ such that $\underline{v}_{t+1} < \pi(x;\, \mathcal{D}_t) \leq \underline{v}_t$. By the monotonicity of $\underline{v}_t$ and $\pi(x;\, \mathcal{D}_{t})$ in $t$, we have $\pi(x;\, \mathcal{D}_{t'}) > \underline{v}_{t'+1}$ for all $t' > t$. Similarly, we have $\pi(x;\, \mathcal{D}_{t'}) \leq \underline{v}_t \leq  \underline{v}_{t'+1}$ for all $t'< t$.  These inequalities establish the validity of the binary search algorithm, which solves $O(\log J)$ instances of $\mathsf{P}^{\dagger}_{\textsf{LP}}(x, {\cal D}_t)$.
\end{proof}

By Theorem~\ref{thm:interpolation-binary}, evaluating $\psi_{{\cal F}}$ is computationally extremely cheap once the value problem is solved. Also note that Algorithm~\ref{algo:interpolation-binary} is exact, it returns the value $\psi_{{\cal F}}(x)$ (rather than an approximation of it).

\section{Extensions}
\label{sec:extensions}

In this section we discuss some extensions of our main model.

\paragraph{Non-Monotone Functions}
We can easily drop the requirement $f \in {\cal F}_{\text{Mon}}$, which does not materially change the sorting algorithm (we still do the sorting in decreasing order of $\psi_{{\cal F}}-$values).
This change is reflected by dropping the constraint $s \geq 0$ from $\mathsf{P}$ and the constraint $\xi \geq 0$ from $\mathsf{P}_{\textsf{LP}}(x;\,\mathcal{D}_t)$.

\paragraph{Non-Permutation Invariant Functions}
When $K=N$ in [Prm], admissible functions are not required to satisfy any form of permutation invariance. The set of admissible functions is then just:
\begin{equation}
\label{eq:admissible-non-prm}
{\cal F} = {\cal F}(\mathsf{D}, {\cal R}, L) := {\cal F}_{\text{QCo}} \cap {\cal F}_{\text{Mon}} \cap {\cal F}_{\text{Mjr}}(\mathsf{D}) \cap {\cal F}_{\text{Rnk}}({\cal R}) \cap {\cal F}_{\text{Lip}}(L).
\end{equation}
In this case we have $\Sigma(\Theta) = \Theta$, and we redefine $\widehat{\Theta} := \{ (\theta, \theta') \in \Theta \times \Theta : \theta \ne \theta'\}$ to be the set of all edges in $\Theta$.
The value problem simplifies to:
\begin{eqnarray*}
\min_{v,\,s}\, && \sum_{\theta \in \Theta}v(\theta)\label{prob:value-non-prm-1}\\
{\rm s.t.}\, && v(\theta)+\max\left\{ \langle s(\theta),\,\theta'-\theta\rangle,\,0\right\} \geq v(\theta'), \quad \forall\left(\theta,\,\theta'\right)\in\widehat{\Theta},\label{prob:value-non-prm-2}\\
 && v(\theta) \geq v(\theta'), \quad  \forall\left(\theta,\,\theta'\right)\in {\cal R},\label{prob:value-non-prm-3}\\
 && v(\theta) \geq \hat{v}(\theta),\, s(\theta)\geq0,\,\|s(\theta)\|_{1}\leq L, \quad\forall \theta\in \Theta,\label{prob:value-non-prm-4}
\end{eqnarray*}
and the prediction problem at $\theta$ becomes:
\begin{subequations}\label{eq:descent}
\begin{eqnarray}
\min _{\upsilon, \xi} \, && \upsilon\\
{\rm s.t.} \, && \upsilon+\left\langle \xi, \theta'-\theta\right\rangle \geq v^*(\theta'), \quad \forall \theta' \in \mathcal{D}_t,\\
&& \upsilon \geq v^*(\theta'), \quad \forall(\theta, \theta') \in {\cal R}, \theta' \in \mathcal{D}_t,\\
&& \upsilon \geq \hat{v}(\theta), \xi \geq 0,\left\|\xi\right\|_{1} \leq L.
\end{eqnarray}
\end{subequations}
Problem~\eqref{eq:descent} has only a polynomial number of decision variables and constraints, in contrast to Problem~\eqref{eq:descent-permutationinvariance-multi}. It can already be practically solved in its existing form.

\paragraph{Least Squares}
Mukherjee et al. \cite{mukherjee2024least} propose a least squares regression problem for $\mathsf{D}$:
\begin{equation}
\label{prob:least_squares}
     \min_{f\in {\cal F}_{\text{QCo}}}\sum_{\theta \in \Theta} (f(\theta) - \hat{v}(\theta))^2.
\end{equation}
Our current sorting algorithm does not apply to Problem~\eqref{prob:least_squares}.
Algorithm~\ref{algo:sorting} relies on the fact that the value problem (which minimizes the summation $\min \sum_{\theta \in \Theta} v(\theta)$) can be decomposed into individual subproblems (which just minimize $\min v(\theta)$) for each $\theta$. Problem~\eqref{prob:least_squares} does not have such a decomposition due to the non-separable objective. 
However, we can still apply our interpolation procedure to the solution of Problem~\eqref{prob:least_squares}.
Problem~\eqref{prob:least_squares} is solved in \cite{mukherjee2024least} for the class of piece-wise constant quasiconcave functions (without a Lipschitz condition), by reformulating it as an MIQP.



\section{Numerical Experiments}\label{sec:numerical}

We have conducted numerical experiments to demonstrate the effectiveness of our proposed method. 
In this section, we begin by describing the experimental setting and then we report the test results.
We want to estimate an unknown production function, which represents the relationship between the final output and the inputs in an economy. Suppose the true production function has the Cobb–Douglas form: 
\begin{equation}
\label{eq:cobb-douglas}
    f(x) = \gamma x_1^{\alpha_1} x_2^{\alpha_2}, \quad (x_1, x_2) \in [x_{\min},x_{\max}] \times [x_{\min},x_{\max}],
\end{equation}
where $f(x)$ represents the total production as a function of $x = (x_1, x_2)$, $x_1$ represents the first input (e.g., capital), $x_2$ represents the second input (e.g., labor), $0 \leq x_{\min} < x_{\max} < \infty$ are lower and upper bounds on each input, and $\gamma > 0$. In the original Cobb–Douglas production function, the output elasticities $\alpha_1,\alpha_2 >0$ must belong to $(0,1)$ and sum to one. However, with knowledge externality, we often have  $\alpha_1+\alpha_2 >1$, which represents an increasing marginal return (see, e.g., \cite{romer1990endogenous}). When $\alpha_1+\alpha_2 >1$, the production function is monotone, nonnegative, and quasiconcave. We would like to find the most conservative estimate (QCoE) of the production function that satisfies these basic properties, with additional partial information. Here we focus on the non-permutation invariant case and do not enforce [Prm]. Under [Prm], the scalability of our approach is even better compared to the MILP reformulation.

Next, we specify how the set of partial information is generated: 
\begin{itemize}
    \item We randomly select an input $x_j = (x_{j1}, x_{j2}) \in \mathbb R^2$ uniformly from the square $[x_{\min},x_{\max}] \times [x_{\min},x_{\max}]$. We know the true function value at $x_j$ must satisfy $f(x_j) \geq \hat{v}(x_j)$. We have $n_{\textsf{mjr}} \geq 1$ such samples.
    
    \item We randomly select two inputs $x_j$ and $x'_j$ from the samples $\{x_1, x_2, \dots, x_{n_{\textsf{mjr}}}\}$. Based on the true function values, we set whether $f(x_j) \geq f(x'_j)$ or $f(x_j) \leq f(x'_j)$. We have $n_{\textsf{rnk}} \geq 1$ such comparisons, which represents our knowledge about the relative value of the function at two different inputs. 

    \item In addition, we know that the production function is $L-$Lipschitz. In reality, it may be difficult to specify $L$ for an unknown function. We discuss how to determine $L$ in Section~\ref{subsec:Lip}.
    
\end{itemize}

In all the experiments, unless otherwise specified, we set $\gamma = 0.1$, $\alpha_1 = 0.5$, $\alpha_2 = 1.5$, $x_{\min} = 0.1$, $x_{\max} = 10$, and $L = 2.0$. 

\subsection{Scalability}\label{subsec:Scalability}
Here we report the scalability of our proposed algorithms for increasing $n_{\textsf{mjr}}$ and $n_{\textsf{rnk}}$. The benchmark algorithm solves the MILP reformulation of Problem~\eqref{prob:value} (the value problem) and Problem~\eqref{prob:interpolation} (the interpolation problem) using the Big-M method, while our procedure is based on solving LPs. One drawback of the MILP reformulation is that it becomes prohibitive to solve when the number of inputs increases. Note that the dimension of the value and interpolation problems mainly depends on the number of sampled inputs $J$. Therefore, in this experiment, we increase the number of sampled inputs $n_{\textsf{mjr}}$ and set $n_{\textsf{rnk}} = n_{\textsf{mjr}}$. This setting reflects the reality of getting more information on the production function as our dataset grows. This experiment is run using Python 3.10 with the COIN-OR Branch and Cut solver (CBC) on a Macintosh platform with an M1 chip and 16GB RAM.

\begin{figure}[h]
    \centering
    \includegraphics[width=0.7\linewidth]{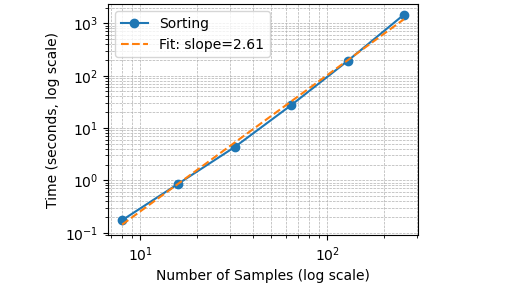}
    \caption{Scalability of Sorting Algorithm: Time vs. Number of Samples in Log Scale}
    \label{fig:scalability-fit}
\end{figure}

Table~\ref{table:scalability-sorting} suggests that our sorting algorithm is  scalable with respect to increasing $n_{\textsf{mjr}}$. However, the MILP reformulation suffers from exponential growth in its complexity as the running time exceeds 30 minutes when $n_{\textsf{mjr}} \geq 128$. In Figure~\ref{fig:scalability-fit}, we plot the running time of the sorting algorithm in log scale and find the best linear fit. We observe that its running time scales on the order of 2.61 with $n_{\textsf{mjr}}$, which matches our complexity result in Theorem~\ref{thm:sorting} as the algorithm solves $O(J^2)$ LPs. The order is slightly larger than 2 because the dimension of the LPs also increases with $J$.

\begin{table}[ht]
\begin{centering}

\begin{tabular}{cccccccccccc}
\hline
 & &  \multicolumn{7}{c}{Number of Sampled Inputs $n_{\textsf{mjr}}$}\tabularnewline
 \cline{3-9}

 & Method &  4 &  8 &  16 &  32 & 64 & 128 & 256
\tabularnewline
\hline 
& Sorting & 0.11 & 0.17 & 0.86 & 4.33 & 27.27 & 190.44 & 1464.41\\

& MILP & 0.05 & 0.20 & 0.81 & 5.15 & 107.42 & --- & ---\\
\hline 
\end{tabular}
\par\end{centering}
\caption{Running time of the sorting algorithm compared to MILP reformulation (in seconds).}
\footnotesize{Note: ``---'' denotes timeout (exceeded 30 minutes).}
\label{table:scalability-sorting}
\end{table}

In Table~\ref{table:scalability-binary}, we report the scalability of both approaches for solving the interpolation problem. We note that the number of decision variables only increases linearly with the number of samples in the interpolation problem.
Though both algorithms solve this problem efficiently, our binary search algorithm becomes relatively more efficient as the sample size increases. Compared with the results in Table~\ref{table:scalability-sorting}, we observe that solving the value problem represents the major bottleneck in computing the QCoE, while interpolation can be done relatively  quickly. 

\begin{table}[ht]
\begin{centering}

\begin{tabular}{ccccccccccc}
\hline
 & &  \multicolumn{7}{c}{Number of Sampled Inputs $n_{\textsf{mjr}}$}\tabularnewline
 \cline{3-9}

 & Method &  4 &  8 &  16 &  32 & 64 & 128 & 256
\tabularnewline
\hline

& Binary Search & 0.02 & 0.02 & 0.04 & 0.06 & 0.11 & 0.22 & 0.56 \\

& MILP & 0.02 & 0.02 & 0.04 & 0.07 & 0.14 & 0.29 & 0.63\\

\hline 
\end{tabular}
\par\end{centering}
\caption{Running time of the binary search algorithm compared to the MILP reformulation (in seconds)}
\label{table:scalability-binary}
\end{table}

\subsection{Comparison of QCoE and True Function} \label{subsec:Comparison}

In this experiment, we compare the true production function against the QCoE obtained using our method.
We choose $n_{\textsf{mjr}} = 100$ and $n_{\textsf{rnk}} = 200$.
We set lower bounds at the sampled points according to $\hat{v}(x_j) = f(x_j) - B_j$ where $B_j$ is exponentially distributed with mean $0.5$. 
The parameter $B_j$ represents varying accuracy of our knowledge of the true function value at different sampled inputs.

\begin{figure}[htbp]
  \centering

  \begin{subfigure}[t]{0.48\textwidth}
    \centering
    \includegraphics[width=1\linewidth]{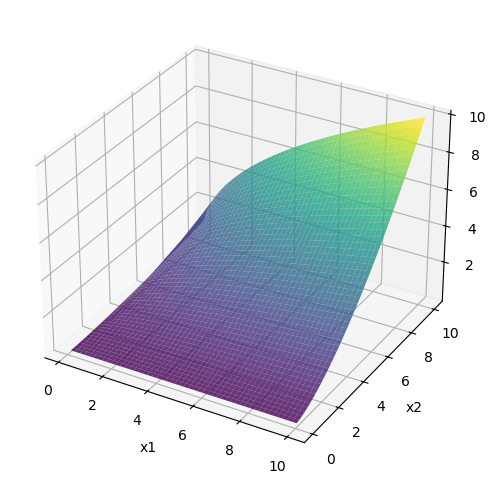}  
    \caption{True Function Surface}                           
    \label{fig:sub1}
  \end{subfigure}
  \hfill
  \begin{subfigure}[t]{0.48\textwidth}
    \centering
    \includegraphics[width=1\linewidth]{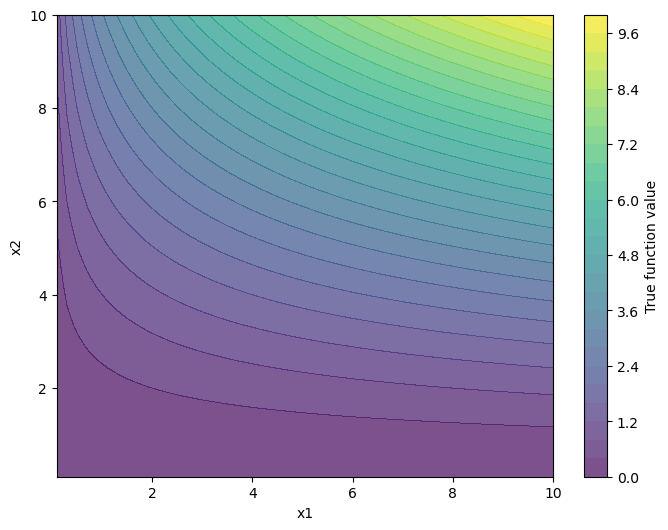}
    \caption{True Function Contour Lines}
    \label{fig:sub2}
  \end{subfigure}

  \vspace{1em}  

  \begin{subfigure}[t]{0.48\textwidth}
    \centering
    \includegraphics[width=1\linewidth]{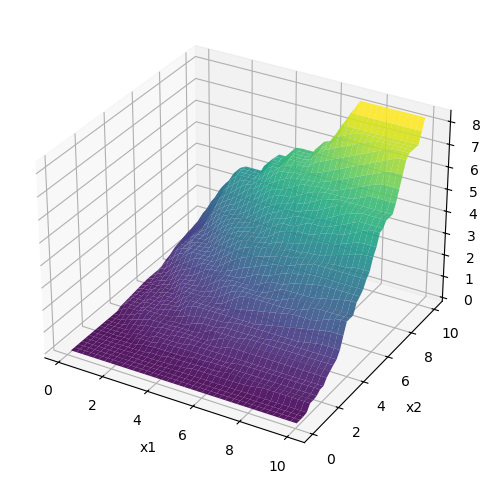}
    \caption{QCoE Surface}
    \label{fig:sub3}
  \end{subfigure}
  \hfill
  \begin{subfigure}[t]{0.48\textwidth}
    \centering
    \includegraphics[width=1\linewidth]{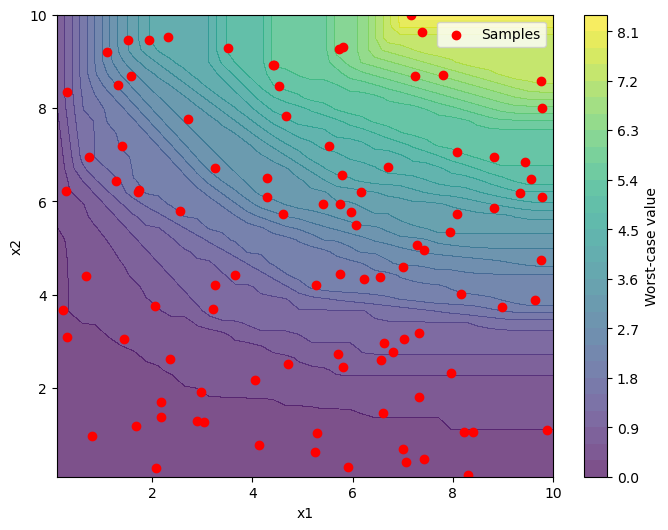}
    \caption{QCoE Contour Lines}
    \label{fig:sub4}
  \end{subfigure}

  \caption{Comparison of True Function and QCoE}
  \label{fig:comparison-true-function-and-QCoE}
\end{figure}

In Figure~\ref{fig:comparison-true-function-and-QCoE}, we plot the function surface and the contour lines of the true production function (top two panels) with the QCoE (bottom two panels).
In the top two panels, we observe that the true production function is quasiconcave with convex upper-level sets. The QCoE is visualized in the bottom two panels. In particular, the QCoE based on the given partial information closely matches the overall shape of the true production function. In addition, we observe that the upper level sets in the contour plot are piece-wise linear and convex, and their vertices coincide with the sampled inputs. 

\subsection{Choice of Lipschitz Constant} \label{subsec:Lip}

\begin{figure}[htbp]
  \centering
  
  \begin{subfigure}[t]{0.24\textwidth}
    \centering
    \includegraphics[width=\linewidth]{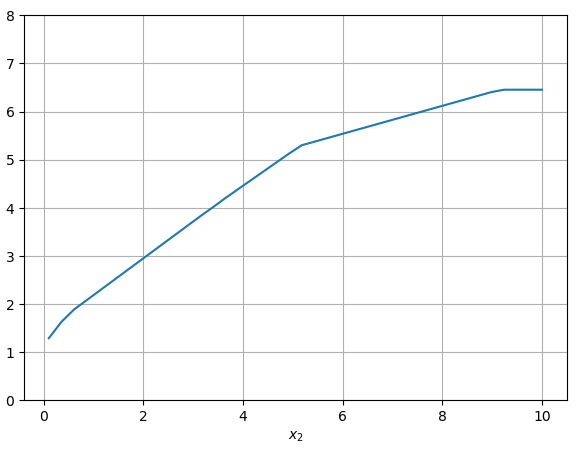}  
    \caption{$L=1.5$}
  \end{subfigure}
  \hfill
  \begin{subfigure}[t]{0.24\textwidth}
    \centering
    \includegraphics[width=\linewidth]{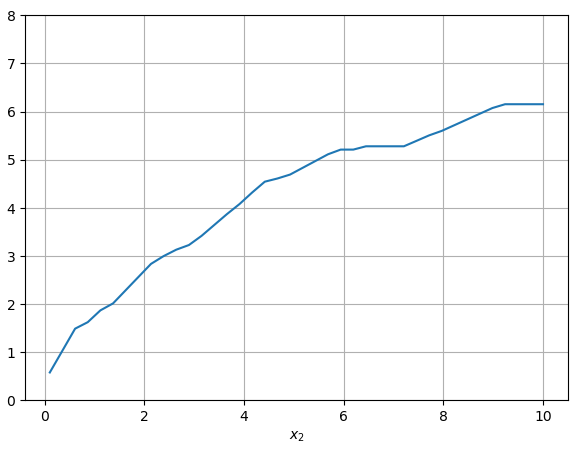}
    \caption{$L=2.0$}
  \end{subfigure}
  \hfill
  \begin{subfigure}[t]{0.24\textwidth}
    \centering
    \includegraphics[width=\linewidth]{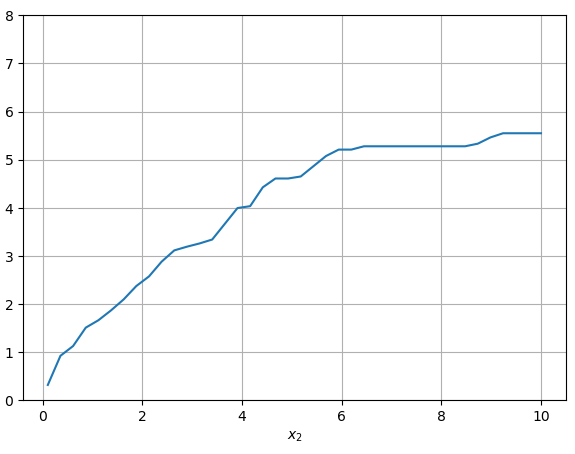}
    \caption{$L=3.0$}
  \end{subfigure}
  \hfill
  \begin{subfigure}[t]{0.24\textwidth}
    \centering
    \includegraphics[width=\linewidth]{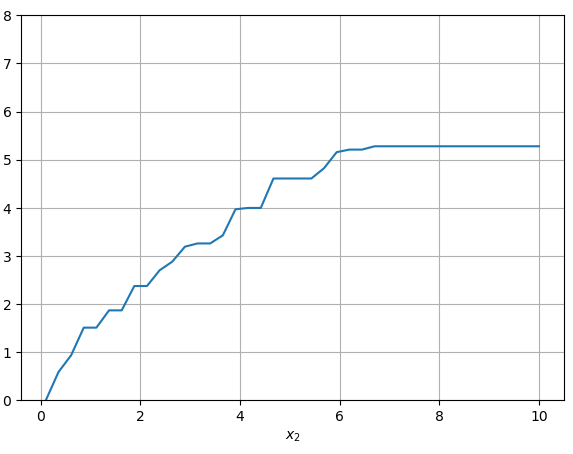}
    \caption{$L=10.0$}
  \end{subfigure}

  \caption{QCoE Value for $x_1 = 8.0$ with Different Choices of $L$}
  \label{fig:different-L}
\end{figure}

Unlike in the PRO setting \cite{haskell2022preference}, the choice of the Lipschitz constant $L$ matters here because we are interested in the actual function values (while \cite{haskell2022preference} just cares about the ordinal values).
However, without knowing the true production function, it might be difficult to specify an appropriate Lipschitz constant.
In Figure~\ref{fig:different-L}, 
we plot the QCoE given different values of $L$. Because the function is two-dimensional, for illustration we slice at $x_1=8.0$ and plot the QCoE value along the $x_2$ axis. As we can see, a larger Lipschitz constant $L$ leads to a more conservative QCoE as the function values are smaller when $L$ is larger. As $L$ continues to grow, the QCoE approaches a piece-wise constant function (see Figure~\ref{fig:different-L}(d)). Therefore, in practice we can start by initially choosing a large $L$ and then gradually reduce $L$ until we obtain a QCoE with reasonable smoothness.

Suppose that we additionally have upper bounds $\bar{v}(x_j)$ on the true function value where $f(x_j) \leq \bar{v}(x_j)$ for all $j \in [n_{\textsf{mjr}}]$. Then, we can continue to reduce $L$ as long as the QCoE at $x_j$ does not violate any of these upper bounds.

\section{Conclusion}\label{sec:conclusion}

This paper has put forward a new method for constructing the QCoE of an unknown quasiconcave function based on limited information. We only know lower bounds on the values of the target function on a set of sample points, and we have some knowledge about its functional properties (monotonicity, Lipschitz continuity, ranking, and permutation invariance).
This task is challenging because the property of quasiconcavity makes the value and interpolation problems non-convex. Furthermore, the property of permutation invariance requires an exponential number of decision variables and constraints in the value problem, and an exponential number of constraints in the interpolation problem.
As our main methodological contribution, we develop a sorting algorithm which can efficiently construct the QCoE by solving the value problem via only a polynomial number of tractable LPs. With the solution of the value problem in hand, we develop a binary search algorithm which can do exact interpolation for points outside the data sample by solving only a logarithmic number of tractable LPs. Our numerical experiments on estimation of a production function, demonstrate the superiority of our approach over the conventional MILP reformulation in terms of efficiency and scalability.

While we have emphasized quasiconcave functions, our approach immediately applies to quasiconvex functions with straightforward modification.
Our approach can also potentially be used in other applications with ambiguous quasiconvex or quasiconcave objectives, or in functional estimation with quasiconvexity/quasiconcavity constraints.
In future research, we can further build on our sorting algorithm to expand the class of ${\cal F}$ for which $\psi_{{\cal F}}$ can be constructed efficiently. We can also connect the present approach with robust optimization for quasiconcave maximization problems under uncertainty.

\bibliographystyle{siamplain}
\bibliography{references.bib}
\end{document}